\documentclass[10pt,conference]{IEEEtran}
\IEEEoverridecommandlockouts
\newcommand{\insertFigure}[2]{
    \begin{figure}[t!]
        \centering
        \includegraphics[width=\linewidth]{fig/#1.pdf}
	\vspace{-5mm}
        \caption{ #2}
	\vspace{-5mm}
        \label{fig:#1}
    \end{figure}
}

\newcommand{\insertWideFigureRatio}[3]{
    \begin{figure*}[ht!]
        \centering
        \includegraphics[width=#3\textwidth]{fig/#1.pdf}
	\vspace{-2mm}
        \caption{ #2}
	\vspace{-3mm}
        \label{fig:#1}
    \end{figure*}
}

\newcommand{\squishlist}{
 \begin{list}{$\bullet$}
  { \setlength{\itemsep}{0pt}
     \setlength{\parsep}{0pt}
     \setlength{\topsep}{3pt}
     \setlength{\partopsep}{0pt}
     \setlength{\leftmargin}{1.5em}
     \setlength{\labelwidth}{1em}
     \setlength{\labelsep}{0.5em} } }

\newcommand{\squishnums}{
 \begin{list}{$\bullets$}
  { \setlength{\itemsep}{0pt}
     \setlength{\parsep}{3pt}
     \setlength{\topsep}{3pt}
     \setlength{\partopsep}{0pt}
     \setlength{\leftmargin}{1.5em}
     \setlength{\labelwidth}{1em}
     \setlength{\labelsep}{0.5em} } }

\newcommand{\squishlisttwo}{
 \begin{list}{$\bullet$}
  { \setlength{\itemsep}{0pt}
     \setlength{\parsep}{0pt}
    \setlength{\topsep}{0pt}
    \setlength{\partopsep}{0pt}
    \setlength{\leftmargin}{2em}
    \setlength{\labelwidth}{1.5em}
    \setlength{\labelsep}{0.5em} } }

\newcommand{\squishend}{
  \end{list}  }

\newcommand{\squishnobullet}{
 \begin{list}{}
  { \setlength{\itemsep}{0pt}
     \setlength{\parsep}{0pt}
     \setlength{\topsep}{3pt}
     \setlength{\partopsep}{0pt}
     \setlength{\leftmargin}{0.5em}
     \setlength{\labelwidth}{1em}
     \setlength{\labelsep}{0.5em} } }
\newcommand{\compiler}{{\textrm{CROSS}}\xspace}
\newcommand*\circled[1]{\tikz[baseline=(char.base)]{
            \node[shape=circle,draw,inner sep=0.5pt] (char) {#1};}}
\newcommand{\+}{{\text{+}}}

\newcommand{\secref}[1]{\S\ref{#1}}
\newcommand{\figref}[1]{Fig.~\ref{#1}}
\newcommand{\tabref}[1]{Tab.~\ref{#1}}

\usepackage{mathptmx} 

\usepackage[normalem]{ulem}
\usepackage{flushend}

\usepackage{pifont}
\usepackage{comment}
\usepackage{xspace}
\usepackage[table]{xcolor}
\usepackage{soul}
\usepackage{makecell}
\usepackage{subcaption}
\usepackage{pifont}
\usepackage{multirow}
\usepackage{algorithm}
\usepackage[noend]{algpseudocode}
\usepackage{fancyvrb}

\usepackage{arydshln}
\usepackage{stmaryrd}
\usepackage{tikz}
\usepackage[english]{babel}

\usepackage{cite}
\usepackage{amsmath,amssymb,amsfonts}
\usepackage{graphicx}
\usepackage{textcomp}
\usepackage{xcolor}
\usepackage[hyphens]{url}
\usepackage{fancyhdr}
\usepackage{hyperref}

\usepackage{listings}
\newcommand{\hpcayear}{2026}

\title{\huge Leveraging ASIC AI Chips for Homomorphic Encryption}

\def\aeopen{}           
\def\aereviewed{}     
\def\aereproduced{} 

\definecolor{colorRevA}{rgb}{0.85, 0.90, 0.98}
\definecolor{colorRevB}{rgb}{0.97, 0.80, 0.796}
\definecolor{colorRevC}{rgb}{0.87, 0.83, 0.90}
\definecolor{colorRevD}{rgb}{0.83, 0.91, 0.83}
\definecolor{colorRevE}{rgb}{1, 0.90, 0.80}
\definecolor{colorRevF}{rgb}{0.77, 0.78, 0.83}
\definecolor{blond}{rgb}{0.98, 0.94, 0.75}

\newif\ifrevisionon
\revisionontrue

\DeclareRobustCommand{\RevC}[1]{#1}
\DeclareRobustCommand{\RevD}[1]{#1}
\DeclareRobustCommand{\RevE}[1]{#1}
\DeclareRobustCommand{\RevF}[1]{#1}

\DeclareRobustCommand{\GPU}[1]{#1}

\author{
\hspace{-4mm}
  \IEEEauthorblockN{{Jianming Tong}\IEEEauthorrefmark{1}\IEEEauthorrefmark{2},
                    {Tianhao Huang}\IEEEauthorrefmark{2},
                    {Jingtian Dang}\IEEEauthorrefmark{1},
                    {Leo de Castro}\IEEEauthorrefmark{2},
                    {Anirudh Itagi}\IEEEauthorrefmark{1}, 
                    {Anupam Golder}\IEEEauthorrefmark{1},
                    {Asra Ali}\IEEEauthorrefmark{3}}
                    
\IEEEauthorblockN{{Jeremy Kun}\IEEEauthorrefmark{3}, 
{Jevin Jiang}\IEEEauthorrefmark{3}, 
                    {Arvind}\IEEEauthorrefmark{2}\IEEEauthorrefmark{5}\thanks{\IEEEauthorrefmark{5}We appreciate Arvind’s technical feedback and strong encouragement. Unfortunately Arvind passed away before this work is published, but we feel extremely honored to have this paper become a part of his strong legacy.}
                    {G. Edward Suh}\IEEEauthorrefmark{4}
                    \thanks{\IEEEauthorrefmark{4}This work was done before this author joined NVIDIA.},
                    {Tushar Krishna}\IEEEauthorrefmark{1}
                    }
\IEEEauthorblockA{\IEEEauthorrefmark{1}Georgia Institute of Technology \IEEEauthorrefmark{2}Massachusetts Institute of Technology \IEEEauthorrefmark{3}Google
\IEEEauthorrefmark{4}NVIDIA/Cornell University}
jianming.tong@gatech.edu, tianhaoh@mit.edu, \{asraa, jkun, jevinjiang\}@google.com, suh@ece.cornell.edu, tushar@ece.gatech.edu
}

\fancypagestyle{camerareadyfirstpage}{%
  \fancyhead{}
  
  \fancyhead[C]{
    \ifdefined\aeopen
    \parbox[][12mm][t]{13.5cm}{\hpcayear{} IEEE International Symposium on High-Performance Computer Architecture (HPCA)}    
    \else
      \ifdefined\aereviewed
      \parbox[][12mm][t]{13.5cm}{\hpcayear{} IEEE International Symposium on High-Performance Computer Architecture (HPCA)}
      \else
      \ifdefined\aereproduced
      \parbox[][12mm][t]{13.5cm}{\hpcayear{} IEEE International Symposium on High-Performance Computer Architecture (HPCA)}
      \else
      \parbox[][0mm][t]{13.5cm}{\hpcayear{} IEEE International Symposium on High-Performance Computer Architecture (HPCA)}
    \fi 
    \fi 
    \fi 
    \ifdefined\aeopen 
      \includegraphics[width=12mm,height=12mm]{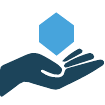}
    \fi 
    \ifdefined\aereviewed
      \includegraphics[width=12mm,height=12mm]{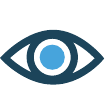}
    \fi 
    \ifdefined\aereproduced
      \includegraphics[width=12mm,height=12mm]{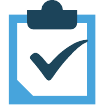}
    \fi
  }
  \fancyfoot[C]{}
}

\begin{document}
\maketitle
\thispagestyle{camerareadyfirstpage}
\pagestyle{empty}

\newcommand{\hpcaheight}{0mm}
\ifdefined\eaopen
\renewcommand{\hpcaheight}{12mm}
\fi


\newif\ifcommenton
\commentontrue

\ifcommenton
\newcommand{\TODO}[1]{\textcolor{red}{[TODO] #1}}
\newcommand{\JT}[1]{{\color{brown}\bfseries [Jianming: #1]}}
\newcommand{\AI}[1]{{\color{blue}\bfseries [Anirudh: #1]}}
\newcommand{\PC}[1]{{\color{blue}\bfseries [Prasanth: #1]}}
\newcommand{\TK}[1]{{\color{violet}\bfseries [TK: #1]}}
\newcommand{\GJ}[1]{{\color{blue}\bfseries [GJ: #1]}}
\newcommand{\THH}[1]{{\color{teal} [Tianhao: #1]}}
\newcommand{\fixme}[1]{{{\color{blue} #1}}}
\else
\newcommand{\AI}[1]{}
\newcommand{\PC}[1]{}
\newcommand{\JT}[1]{}
\newcommand{\TK}[1]{}
\newcommand{\THH}[1]{}
\newcommand{\GJ}[1]{}
\newcommand{\fixme}[1]{}
\fi


\begin{abstract}
\label{sec:abstract}
Homomorphic Encryption (HE) provides strong data privacy for cloud services but at the cost of prohibitive computational overhead. While GPUs have emerged as a practical platform for accelerating HE, there remains an order-of-magnitude energy-efficiency gap compared to specialized (but expensive) HE ASICs. This paper explores an alternate direction: leveraging existing AI accelerators, like Google's TPUs with coarse-grained compute and memory architectures, to offer a path toward ASIC-level energy efficiency for HE.

However, this architectural paradigm creates a fundamental mismatch with SoTA HE algorithms designed for GPUs. These algorithms rely heavily on: (1) high-precision (32-bit) integer arithmetic to now run on a TPU's low-throughput vector unit, leaving its high-throughput low-precision (8-bit) matrix engine (MXU) idle, and (2) fine-grained data permutations that are inefficient on the TPU's coarse-grained memory subsystem. Consequently, porting GPU-optimized HE libraries to TPUs results in severe resource under-utilization and performance degradation.

To tackle above challenges, we introduce \textbf{\compiler}, a compiler framework that systematically transforms HE workloads to align with the TPU's architecture. CROSS makes two key contributions: (1) \textit{Basis-Aligned Transformation (BAT)}, a novel technique that converts high-precision modular arithmetic into dense, low-precision (INT8) matrix multiplications, unlocking and improving the utilization of TPU's MXU for HE, and (2) \textit{Memory-Aligned Transformation (MAT)}, which eliminates costly runtime data reordering by embedding reordering into compute kernels through offline parameter transformation.

Our evaluation on a real single-host Google TPU v6e refreshes the SoTA Number Theoretic Transform (NTT) throughput record with up-to \textit{$1.43\times$ throughput improvement} over WarpDrive on a NVIDIA A100. Furthermore, CROSS achieves $451\times$, $7.81\times$, $1.83\times$, $1.31\times$, $1.86\times$, and $1.15\times$ higher throughput per watt than OpenFHE, WarpDrive, FIDESlib, FAB, HEAP, and Cheddar, respectively, establishing AI ASIC as the SotA efficient platform for HE operators. Code: \url{https://github.com/EfficientPPML/CROSS}.
\end{abstract}


\section{Introduction}
\label{sec:introduction}

Artificial Intelligence (AI) is driving a new industrial revolution, transforming how we create, exchange, and safeguard information. From chatbots\cite{hurst2024gpt,comanici2025gemini25pushingfrontier}, generative model\cite{sengar2025generative} to AI coders\cite{liang2024large}, human workflows are increasingly translated into digital tokens—a process that \textbf{makes the world effectively tokenized}. This revolution necessitates AI systems that learn and operate on sensitive, individual user data, making robust privacy preservation not merely a feature, but a foundational requirement. 
Homomorphic Encryption (HE) offers a powerful solution by enabling direct computation on encrypted data, as shown in \figref{fig:FHE_overview}. But its practical adoption is hindered by extreme performance overheads, with data and computation costs inflating by orders of magnitude, leading to 1000$\times$ slowdown on a multi-core CPU~\cite{f1, CraterLake,mouchet2020lattigo,openfhe}.

Dedicated ASICs~\cite{putra2023strix, shivdikar2023gme, agrawal2023mad, CraterLake, sharp23, BTS, Poseidon_HPCA23, ARK_MICRO22, Choco_taco_ASPLOS22, ASIC_Poly_Mul, cinnamon_siddharth} have been proposed to tackle this challenge, but their high design and fabrication costs present a significant barrier to widespread deployment. Consequently, commodity hardware like GPUs~\cite{fan2022tensorfhe, dathathri2018chet, GPU_WhitePaper_mention_21, FxHENN_HPCA23,IntelHEXL,kim2024cheddar,li2025catgpuacceleratedfheframework, phantom_fhe, agullódomingo2025fideslibfullyfledgedopensourcefhe, wang2024chameleonefficientfhescheme,warpdrive, tensorfhe_plus,HE_Booster} and FPGAs~\cite{HEAX20, agrawal2022fab, CHAM_FPGA} have become the de-facto platforms for high-performance HE, achieving SoTA throughput and energy efficiency. \RevC{While still approximately 33$\times$ less energy-efficient than a well-designed HE ASIC} (\tabref{tab:perf_cmp})\RevC{, the lack of commercial HE chips and the widespread availability of TPU-like AI accelerators make our approach highly compelling.} \RevD{Such a gap is largely attributed to control overheads, mismatching HE computation to the compute pipeline in hardware, extreme data movement, and poor data reuse}~\cite{shivdikar2023gme,SoK_FHE_Accelerator,CraterLake}. \label{q:rd_c2}

\begin{figure}[!t]
    \centering
    \includegraphics[width=\columnwidth]{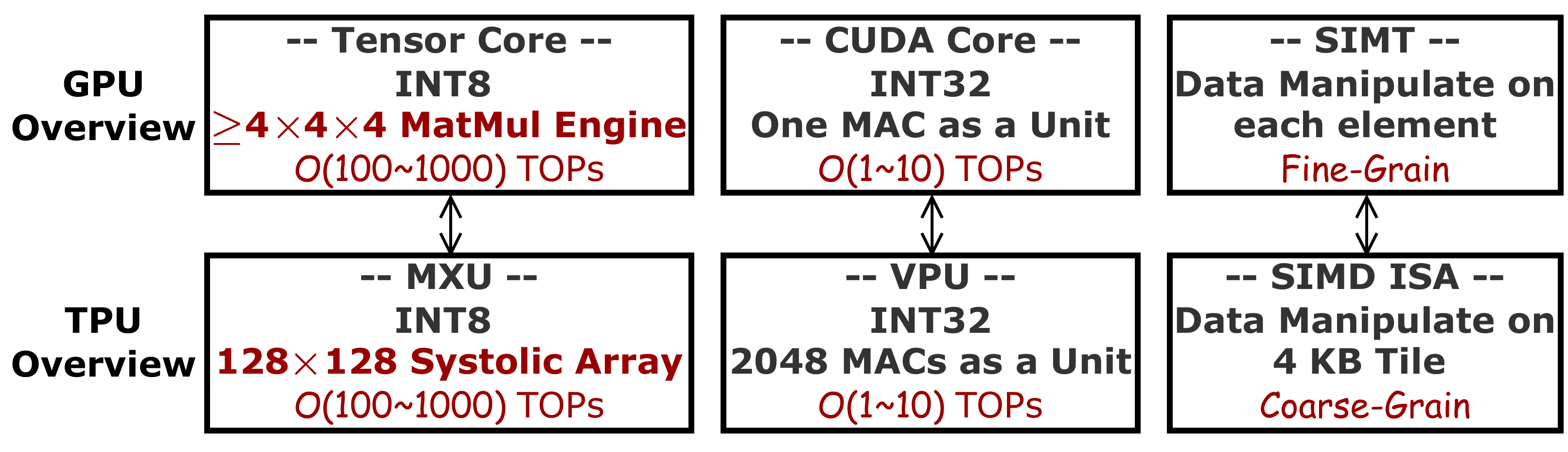}
    \vspace{-6mm}
    \caption{TPU's compute/memory granularity is $>$ GPU.}
    \label{fig:GPU_TPU_diff}
    \vspace{-4mm}
\end{figure}

This paper identifies a compelling alternative: leveraging existing ASIC AI accelerators, such as Google's TPUs, for HE operators to achieve better energy efficiency (performance per watt), because TPU offer \textit{functionally equivalent hardware components as GPUs but employ coarse-grained control to amortize the control overheads} (\figref{fig:GPU_TPU_diff}). However, this coarse-grained design creates a fundamental mismatch with HE algorithms that enables GPU to achieve SoTA throughput. This is because they require high-precision compute and fine-grained data manipulation for each individual data element, and both are inefficient for TPU. Specifically,

\insertFigure{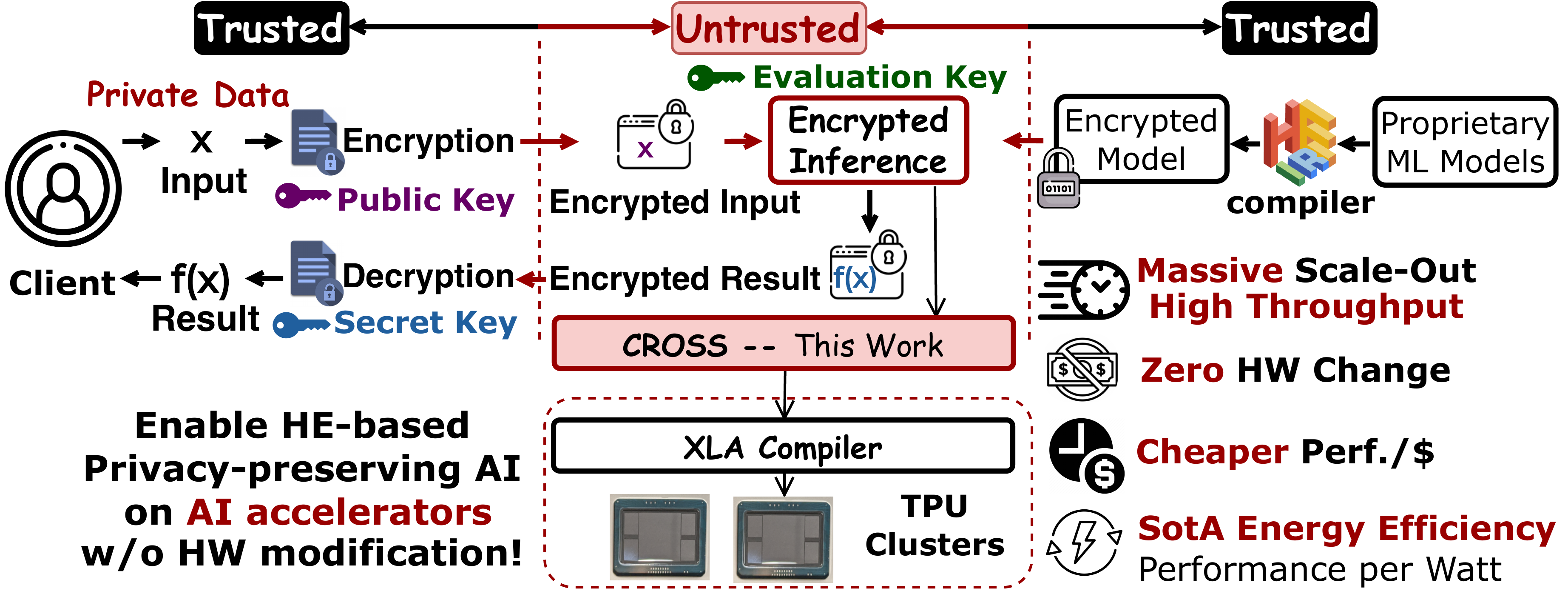}{CROSS enables direct computation on encrypted data to enable privacy-preserving model serving on AI ASICs.}

\textit{\textbf{Arithmetic Mismatch Challenge:}} SoTA GPU HE libraries (1) either have an inefficient precision lowering algorithm to create a sparse matrix with redundant zeros (\ding{182} in \figref{fig:CROSS_overview} with details in \figref{fig:CROSS_Scalar_Flow}), wasting compute and memory and reducing effective compute utilization of matrix multiplication unit (MXU), (2) or rely on 32-bit integer operations~\cite{kim2024cheddar} that are available in TPU's low-throughput Vector Processing Unit (VPU), leaving its powerful, high-throughput low-precision 8-bit MXU fully idle (\ding{183} in Fig.~\ref{fig:CROSS_overview}).

\textit{\textbf{Memory Manipulation Granularity Challenge:}} 
HE kernels like the Number Theoretic Transform (NTT) require frequent, fine-grained data shuffling and transposing. These are prohibitively slow on TPUs, because individual data elements need to be fit into large, coarse-grained, and SIMD-controlled (8, 128) 32-bit registers (4 KB VReg) for achieving desired data manipulation. This reduces effective tile utilization (\ding{184}).

To tackle above divergence, we propose \textit{\textbf{\compiler}}\footnote{\underline{C}ompiling \underline{R}eal-time \underline{O}nline \underline{S}ecure \underline{S}ervice on ASIC AI Accelerators}, a compiler framework that (1) refactors SoTA HE operators into arithmetic that leverage high-throughput MXU with better utilization, and (2) embeds transpose and shuffling into computation in compile time, eliminating runtime data reordering and ensuring a layout-invariant execution. CROSS enables AI accelerators to achieve SoTA energy efficiency (performance per watt) with two architecturally universal optimizations:

\begin{figure}[!t]
    \centering
    \includegraphics[width=\columnwidth]{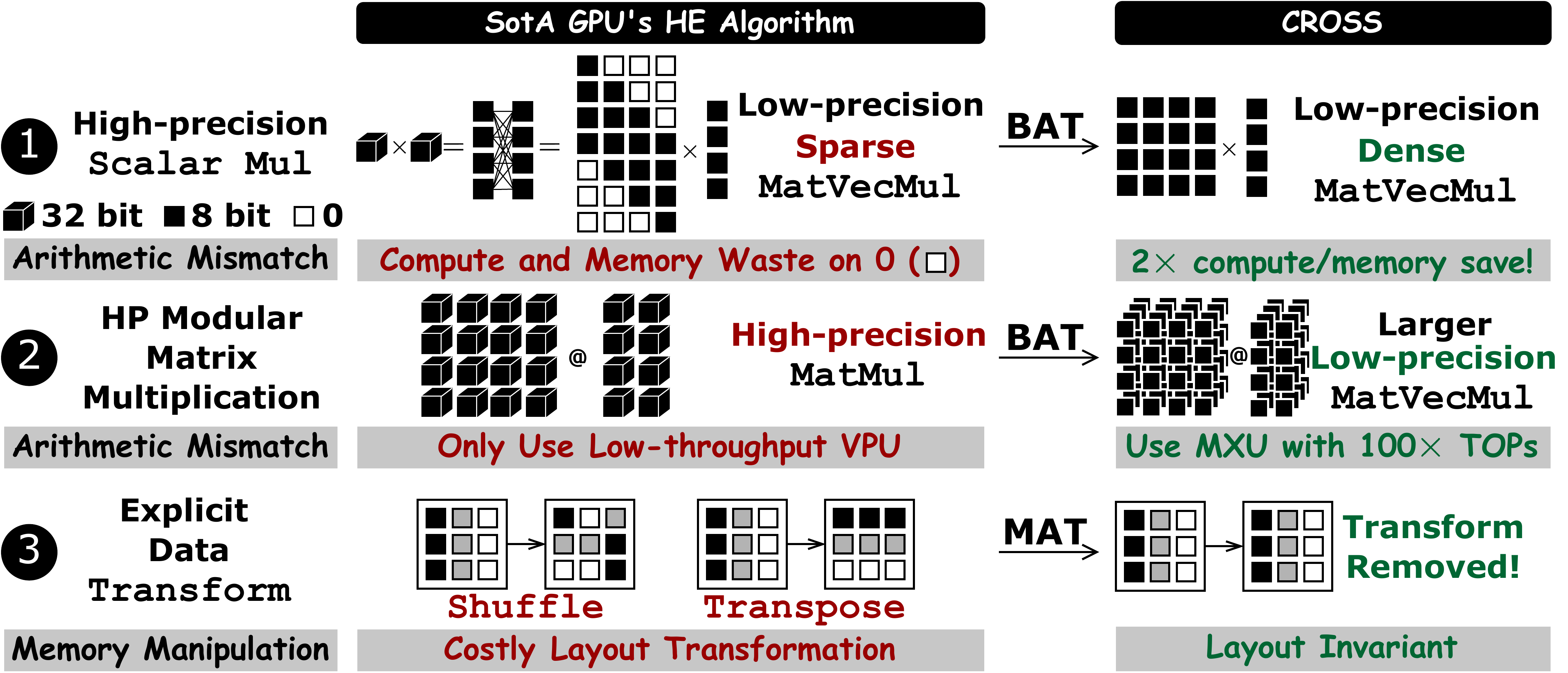}
    \vspace{-6mm}
    \caption{\compiler aims at (1) eliminating compute redundancy, (2) leveraging powerful MXU for throughput improvement, and (3) removing explicit memory costs for better efficiency.}
    \label{fig:CROSS_overview}
    \vspace{-6mm}
\end{figure}

\textbf{\textit{Basis Aligned Transformation (BAT)}} reformulates the sparse left matrix (\ding{182}) of preknown parameters (e.g., twiddle factors in NTT and precomputed primes in Basis Conversion) into a dense half-size low-precision matrix vector multiplication, reducing redundant work and memory waste zeros and increasing compute utilization. Further, BAT converts high-precision (e.g., 32-bit) modular matrix multiplication into dense low-precision (8-bit) matrix multiplication (\ding{183}). 

\textbf{\textit{Memory Aligned Transformation (MAT)}} eliminates expensive runtime data reordering by embedding layout changes, e.g., transpose and shuffling in NTT, directly into the computation. This is achieved by representing reordering as transformation matrix and applies it to pre-known parameters offline (e.g., twiddle factors in NTT), thus creating a ``layout-invariant" kernel with no explicit memory overhead (\ding{184}).

To the best of our knowledge, \compiler represents the first work to investigate deployment of HE operators on AI accelerators, introducing a new paradigm of AI/HE co-acceleration on the same hardware substrate. \compiler shows that AI accelerators without any modifications can indeed be used to accelerate HE operators and achieve SoTA efficiency. 

Our evaluation on a real single-host Google TPUv6e shows that we can 1.2/13$\times$ more throughput of NTT over WarpDrive / TensorFHE+ on NVIDIA-A100, which refreshes the SoTA record of NTT throughput in practically available devices. Furthermore, CROSS achieves $451\times$, $7.81\times$, $1.83\times$, $1.31\times$, $1.86\times$, and $1.15\times$ higher throughput per watt on HE operators than OpenFHE, WarpDrive, FIDESlib, FAB, HEAP, and Cheddar, when scaling TPUv6e to consume roughly the same power as their platforms. These results establish AI accelerators (1) the SoTA energy efficient solution among commodity hardware like GPUs and FPGAs, and (2) a promising platform to investigate for privacy-preserving computation. 

The primary performance gap with HE ASICs (3-33{$\times$}) stems from (1) the absence of low-cost data shuffling engine, (2) no hardware support of carefully selected moduli, and (3) less overall memory/compute, as discussed in \secref{sec:gap_analysis}.


We make the following contributions in this paper:

$\bullet$ A systematic characterization of the architectural mismatches and inefficiencies that arise when running SoTA GPU-based HE algorithms on AI accelerators like Google TPU.

$\bullet$ Basis-Aligned Transformation (BAT), a novel method to map high-precision modular integer arithmetic to low-precision matrix multiplication engines, enabling high-throughput low-precision matrix multiplication engine in  AI accelerators to be used efficiently for HE operators.

$\bullet$ Memory-Aligned Transformation (MAT), a technique to create layout-invariant HE operators that eliminate explicit data transpose and reordering overhead by embedding these operations into computation.

$\bullet$ A comprehensive evaluation demonstrating that CROSS on TPUs achieves SoTA NTT throughput and SoTA energy efficiency for critical HE operators compared to highly optimized CPU, GPU, and FPGA implementations.

\section{Background and Motivation}
\label{sec:motivation}
In this section, we introduce the background for Homomorphic Encryption (HE) and discuss key potential of AI accelerators for HE acceleration.

\vspace{-1mm}
\subsection{Homomorphic Encryption Background}
HE is a specialized form of public-key encryption that enables computations to be performed directly on encrypted data without revealing the underlying plaintext. In an HE system, the client exclusively holds a private decryption key, while a public encryption key is made available for data encryption. Additionally, an evaluation key is provided to the cloud or computing service to facilitate computations on the encrypted data. In HE, direct computation on the encrypted data will apply computation on the underlying messages, such that the entire computation is secured in the privacy-preserving manner as shown in \figref{fig:FHE_overview}.

\subsubsection{\textbf{Terminology and Data Representation}}
The security of HE schemes is based on the hardness of Ring Learning With Errors (RLWE) problem~\cite{Lyu12}. This is a problem over a polynomial ring $R_Q := \mathbb{Z}_Q\left[x\right]\text{/}(x^N\+1)$, where $N$ is a power of two. An element in $R_Q$ is a polynomial of the form $a(x) = \Sigma_{j=0}^{N-1} a_{j}\cdot x^{j}$, where each coefficient $a_j$ is an integer in $\left[0, Q-1 \right]$ and the polynomial is reduced by $x^N\+1$. 

\subsubsection{\textbf{Parameter Determination}} (degree $N$, ciphertext modulus bitwidth $\log_2Q$) are design choices, which determine the security level for a given error standard deviation. Practical applications typically require 128-bit security level, which comes in various choices from ($2^{10}$, 29)~\cite{standard} to ($2^{17}$, 2200$\+$)~\cite{shivdikar2023gme}. A larger coefficient modulus ($\log_2Q$) allows more computation on a ciphertext before bootstrapping is required, but the degree $N$ of the polynomial modulus must grow with $\log_2Q$ in order to maintain security. Larger $Q$ and higher degree both lead to longer computational latency. Therefore, the minimal (degree, $\log_2Q$) that satisfies the required computation is often selected to minimize overheads when bootstrapping is not required.

\begin{table}[!t]\centering
\caption{Notations}
\vspace{-2mm}
\label{tab:Notation}
\scriptsize
\resizebox{\columnwidth}{!}{
\begin{tabular}{ccc}\hline
Term & Conditions & Meaning (example value)\\\hline
\begin{math}N\end{math} & Power of two &Polynomial degree (\begin{math}2^{16}\end{math})\\
\begin{math}Q\end{math} &  &Ciphertext modulus (\begin{math}1728\end{math} bits) \\
\begin{math}q_i\end{math} & coprime &RNS base \begin{math}Q =\Pi_{(i=0)}^{(L-1)} q_i\end{math} (\begin{math}28\end{math} bits)\\
\begin{math}L\end{math} & \begin{math}\log_2\left(Q\right)\end{math}/\begin{math}\log_2\left(q_i\right)\end{math} &The number of limbs (\begin{math}\lceil 1728\text{/}28 \rceil\end{math} limbs) \\
\begin{math}L'\end{math} &  & number of limbs with auxiliary modulus \\
\begin{math}\log_2q\end{math} & & RNS bases have \begin{math}\log_2q\end{math} bits, noted as $\log_2q$ \\
\begin{math}bp\end{math} & &  $\underline{B}$it $\underline{P}$recision of MAC in hardware \\
\begin{math}\omega_n\end{math} &  & A primitive \begin{math}n\end{math}-th root of unity \\
\begin{math}\mathbf{a},\mathbf{b},\mathbf{z}\end{math} &  & Coefficient vector \\
\begin{math}dnum\end{math} &  & Number of digits in the switching key \\
\begin{math}K\end{math}  &  & Number of 8-bit chunks in high-precision scalar  \\
\begin{math}\mathcal{B}\end{math} & \begin{math}\mathcal{B}={q_0, \cdots, q_{L-1}}\end{math} & A set of RNS bases.  \\
\hline
\end{tabular}}
\vspace{-7mm}
\end{table}

\subsubsection{\textbf{Residue Number System (RNS)}}
\label{sec:cross_mod_mul}
Once ($N$, $\log_2Q$) is being made, each raw data will be encoded and encrypted into a ciphertext with a pair of polynomials of degree $N$. Each coefficient would be high-precision, often thousands of bits~\cite{standard, shivdikar2023gme}, which are not natively supported by the 32/64-bit  micro-architectures in CPUs or GPUs. Naively, mapping high-precision data to low-precision computation unit requires two steps: (1) breaking high-precision coefficients into low-precision chunks supported by computation units and (2) executing multiplicative operations across all pairs of chunks from two coefficients. This segmentation and \textit{chunkwise multiplication} incur \textit{quadratic} computational overheads.

To reduce such quadratic pair-wise computation costs, the Chinese Remainder Theorem (CRT) allows us to construct a set of coprime RNS basis $\{q_0, \cdots, q_{L-1}\}$, where $Q = \Pi_{i=0}^{L-1} q_i$. Under CRT, each high-precision coefficient $a_n, n \in \left[0, N-1\right] $ of a polynomial in $R_Q$ is represented as residues of a sequence of $L$ smaller moduli $\{a_n \text{ mod } q_i\}$ for $i \in \left[0, L-1\right] $. These $L$ obtained polynomials with low-precision coefficients are referred to as limbs, noted as ($limb_i$), $i \in \left[0, L-1\right] $.
The isomorphism $R_Q \cong R_{q_0} \otimes \ldots \otimes R_{q_{L-1}}$ allows addition and multiplication to be performed ``limb-wise" over elements of $R_Q$. 
Thus, a limb of one polynomial multiplies only with its counterpart limb of another polynomial, reducing quadratic computational overhead down to linear. In the post-CRT ciphertext, different limbs get processed independently, facilitating limb-level parallelism. A summary of the notations is listed in \tabref{tab:Notation}.

However, RNS cannot directly reduce high-precision data to an arbitrary low precision due to inherent constraints. Specifically, RNS requires a set of moduli that (1) multiply to the original modulus $ Q $, and (2) are pairwise co-prime. On typical AI accelerators~\cite{TPUv2,tpuv4i}, the lowest supported precision is 8-bit integer arithmetic. To reduce 2000-bit data to 8 bits, $ 2000/8 = 250 $ primes are needed. However, it is infeasible to find 250 co-prime integers within the 8-bit range $[0, 256)$. Consequently, precision reduction to 8-bit arithmetic involves two stages:

\noindent (1) \textit{Linear Precision Reduction}: RNS lowers high precision to an intermediate precision (e.g. 32-bit) with linear complexity.
\noindent (2) \textit{Quadratic Precision Reduction}: A secondary reduction further lowers the intermediate precision to 8-bit arithmetic, incurring quadratic complexity, \textit{the costs optimized by \compiler.}

\subsubsection{\textbf{Compute and Memory Analysis}} Above RNS based encoding and encryption introduce about $200\times$ data expansion of the original message, making HE operators memory-bound for devices with small on-chip memory (\textit{memory overhead})~\cite{decastro2021does, jung2021over}. Further, evaluating encrypted data also introduces extra computational complexity, i.e. raw multiplication becomes multiplication of multiple high-degree polynomial rings in HE (\textit{computation overhead}). Both overheads lead to significantly long wall-clock time latency of HE-based privacy-preserving serving. Fortunately, such significantly high compute and memory exhibit inherent independence across ciphertexts, polynomial degree ($N$) and limbs ($L$) etc. Such independence brings high parallelism (eg. up-to $N\cdot L\approx2^{22}$ for a single ciphertext multiplication), offering the possibility of hardware acceleration with sea of compute to reduce latency.

\vspace{-1.5mm}
\subsection{Potential of AI Accelerators for HE Workloads}

\insertFigure{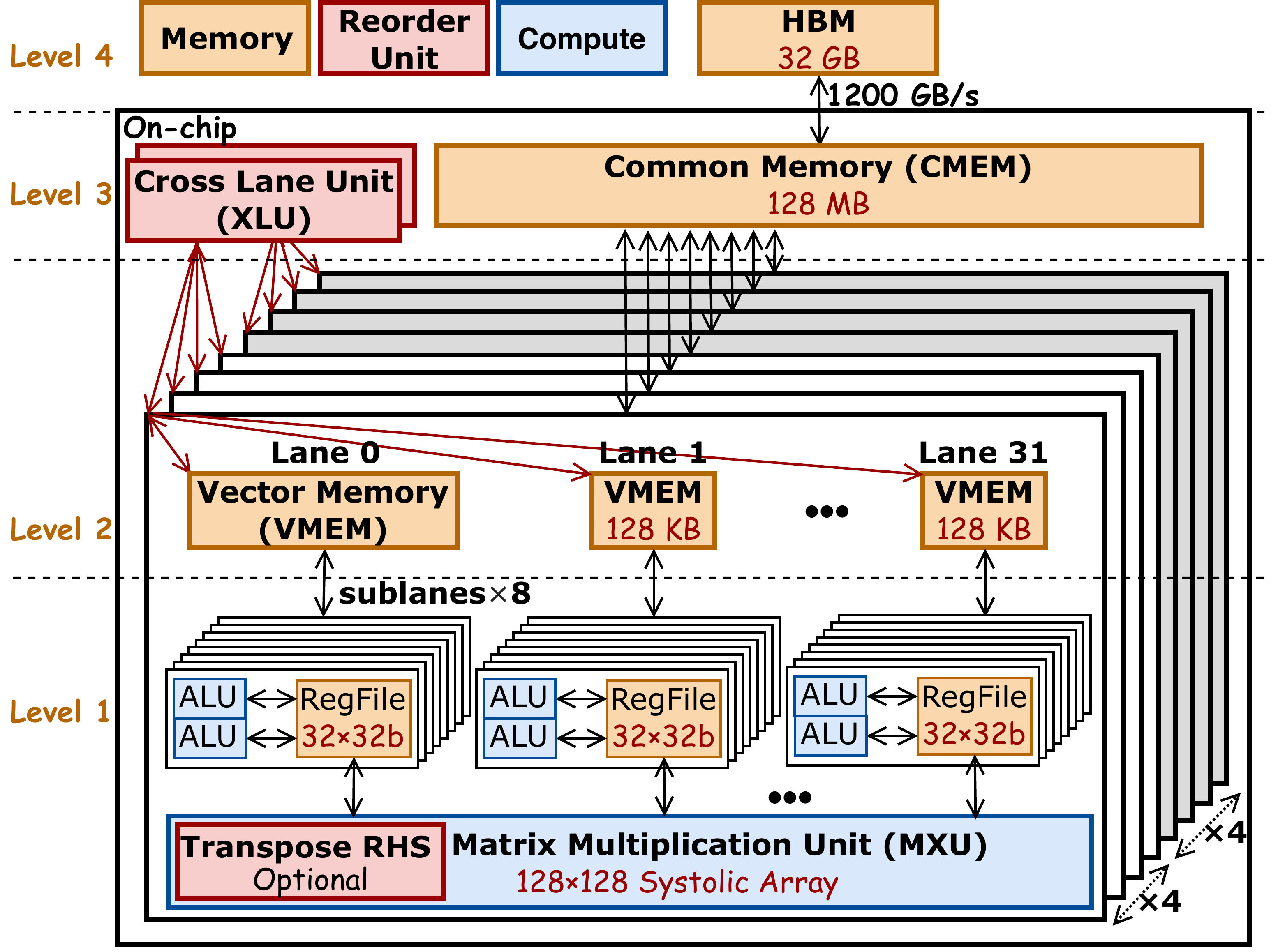}{Overview of TPUv4 architecture based on public information~\cite{tpuv4i,TPUv2,jouppi2023tpu,tpu_web_doc}. Four black and gray boxes represent two tensor cores, separately. Two tensor cores share the same 128 MB common memory (CMEM, removed in newer TPUs) to hold frequently used data. Each tensor core has 4 matrix multiplication units (MXU) and 2048 ALUs in Vector Processing Unit (VPU) organized as 128 SIMD lanes. Each lane consists of 8 SIMD sublanes with 128 KB vector memory (VMEM), each sublane has 2 dual-issue ALUs and 128 B local register file. Such two level of SIMDs force a group of (8, 128) 32-bit registers, termed as VReg, to be operated in the lock step. Each MXU features a $128{\times}128$ systolic array ($256{\times}256$ for TPUs after v6) for performing matrix multiplication. Each MXU has a local transpose unit to optionally transpose right-hand-side (RHS) input matrix in the pipelined manner to hide transpose latency behind. Data in VMEM of different lanes could get transposed or shuffled or accumulated through the Cross Lane Unit (XLU), which consumes non-hidden layout reordering and reduction latency.}

\begin{figure}[!t]
    \centering
    \includegraphics[width=\columnwidth]{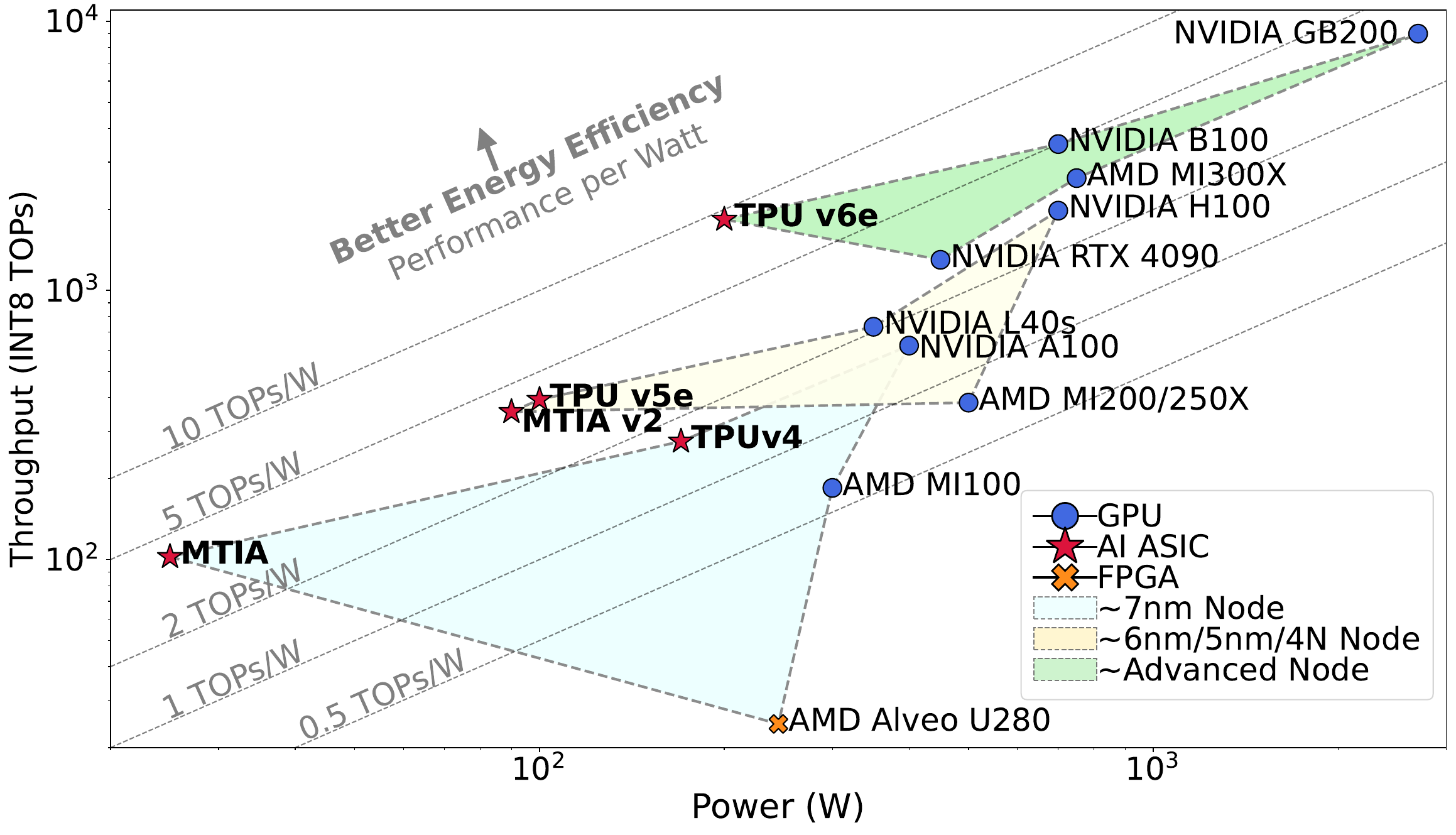}
    \vspace{-7.3mm}
    \caption{AI ASICs deliver better energy efficiency among practical devices using the same technology nodes.}
    \label{fig:energy_efficiency_comparison}
    \vspace{-3mm}
\end{figure}

AI accelerators feature with abundant memory and compute (\figref{fig:tpu_arch_new}), which show great potential for HE acceleration.

$\bullet$ \textbf{Large Compute Array (Parallelism):} \GPU{Each MXU in AI accelerators is $32\times$ larger (e.g., 128$\times$128 in TPUv4) than those in GPUs (typically 4,4,4-Matrix Multiplication~\cite{nvidia-matmul-bg-ug}). The larger size of MXUs increases on-chip data reuse within the two-dimensional computation arrays, enhancing throughput per watt.} Moreover, a sea of 2048 SIMD ALUs, sharing the same VRegs with MXUs and further increasing data reuse. 

$\bullet$ \textbf{Large On-chip Memory:} AI accelerators feature large on-chip memory, e.g. a single Google TPU v4 chip has 160 MB of on-chip memory, including 128 MB CMEM and 32 MB VMEM array in \figref{fig:tpu_arch_new}, \GPU{which is 20$\times$/4$\times$ larger than the AMD MI100 / NVIDIA A100.} This substantial on-chip capacity can accommodate entire ciphertexts to avoid its repeated accesses from off-chip memory, alleviating memory bottleneck inherent in HE workloads for better performance and efficiency.

$\bullet$ \textbf{On-chip Data Management Units:} TPU has specialized Cross Lane Unit (XLU) in \figref{fig:tpu_arch_new}, which could (1) transpose data sitting in on-chip VMEM, (2) shuffle data among VMEMs, and (3) accumulate partial results from VMEMs in different lanes into final results.

\begin{table}[!t]\centering
\caption{Architectural comparison of AI accelerators in \figref{fig:energy_efficiency_comparison}.}\label{tab:arch_high_lvl_diff}
\vspace{-2mm}
\resizebox{0.48\textwidth}{!}{ 
\begin{tabular}{ccccc}\hline
&TPU &GPU &FPGA \\\hline
clock frequency &O(1) GHz &O(1) GHz &O(300) MHz \\
on-chip memory &O(100) MB &O(40) MB &O(40) MB \\
Programming &SIMD &SIMT &Bits-level Reconfiguration \\
INT8 TOPs &O(1000) &O(1000) &O(25) \\
\hline
\end{tabular}}
\vspace{-7mm}
\end{table}

In summary, \textit{large compute array, large on-chip memory and flexible permutation/transpose/reduction engine} enable AI accelerators to achieve better energy efficiency (\textit{performance per watt}, measured by TOPs/watt) over CPU, GPU and FPGAs (see \figref{fig:energy_efficiency_comparison}). However, effectively leveraging these abundant resources to accelerate HE remains a challenge due to the misalignment between compute patterns of SoTA HE algorithms and architectural capabilities of AI accelerators.

The core strategy of \compiler is to systematically remap HE operations to align with the architectural strengths of AI accelerators. \compiler converts high-precision modular arithmetic, the essential computational kernel of HE operators, into dense, low-precision matrix multiplications (\texttt{MatMul}). This transformation is designed to simultaneously embed costly data reordering (transpose and shuffling) directly into the computation, making layout invariant throughput the computation. By doing so, \compiler enables HE workloads to harness the high-throughput matrix engines of AI accelerators like MXU in TPU while sidestepping the performance bottlenecks associated with fine-grained data movement, achieving SoTA energy efficiency on existing AI hardware without any modifications.

With this aim, we first characterize the performance inefficiency when porting SoTA GPU-optimized HE algorithms to AI accelerator using Google's TPU as an example in \secref{sec:misalignment}. To resolve the inefficiency in compute and memory, separately, we propose key two ingredients of \compiler, \textit{Basis Aligned Transformation (BAT)} and \textit{Memory Aligned Transformation (MAT)} in \secref{sec:BAT} and \secref{sec:MAT}. Detailed \compiler evaluation and comparison to SoTAs are analyzed in \secref{sec:evaluation}.

\vspace{-1mm}
\section{Inefficiency of SoTA HE Alg. for AI Accel.}
\label{sec:misalignment}

While GPUs currently deliver SoTA performance for accelerating Homomorphic Encryption (HE) workloads~\cite{kim2024cheddar}, their architectural characteristics differ fundamentally from ASIC-based AI accelerators, such as Google TPUs (\tabref{tab:arch_high_lvl_diff}). This architectural divergence leads to significant inefficiencies when directly porting GPU-optimized HE algorithms onto TPUs. To systematically analyze these inefficiencies, we first abstract SoTA HE acceleration techniques into distinct compilation layers and explicitly define the scope targeted by \compiler. Within this defined scope, we identify and characterize the key sources of performance degradation in GPU-tailored HE kernels on TPUs, motivating a novel compilation approach tailored specifically for efficient HE on AI accelerators.

\begin{figure*}[!t]
    \centering
    \includegraphics[width=\textwidth]{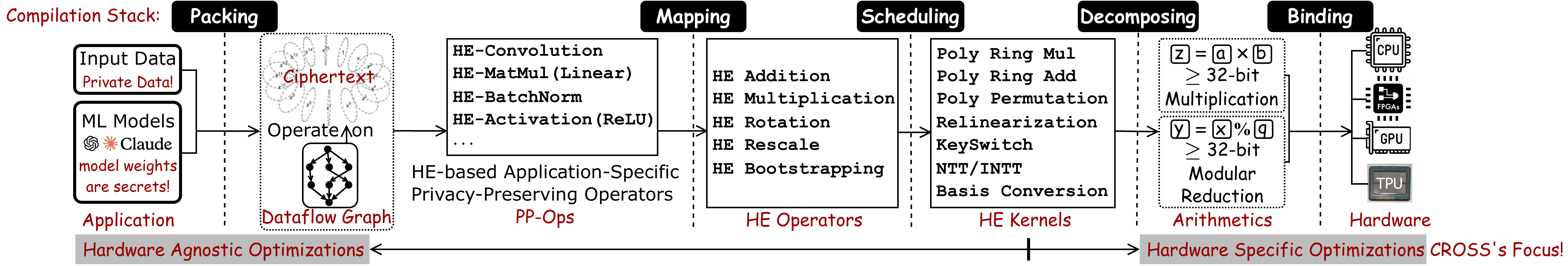}
    \vspace{-6mm}
    \caption{Abstract compilation layers for accelerating HE-based privacy-preserving applications.}
    \label{fig:he_overview}
    \vspace{-6.5mm}
\end{figure*}

\subsection{Optimizations Overview of HE Acceleration Stack}
HE acceleration techniques, particularly for CKKS encryption, could be formally categorized into five distinct layers, including Packing, Mapping, Scheduling, Decomposing and Binding, as illustrated in \figref{fig:he_overview}.

$\bullet$ \textbf{Packing} defines how data is organized within ciphertext slots. Specifically, a CKKS ciphertext operates like a vector of SIMD units, encoding multiple data per ciphertext and enforcing lock-step operator for all encoded data. Thus, operators in original application initially designed for element-wise computations must be transformed into SIMD-compatible, HE-specific Privacy-Preserving Operators (PP-Ops).

$\bullet$ \textbf{Mapping} translates PP-Ops into sequences of fundamental HE operators (see \figref{fig:he_overview}). Optimal mapping seeks maximum arithmetic intensity, data reuse, and parallelism while minimizing computational and memory overhead to reduce latency. Application-specific optimizations typically reside here~\cite{lee_hecate, lee2023elasm, cheon2024dacapo, lee2024performance, Gazelle, orion_nyu}. Each HE operator consists of multiple HE kernels.

$\bullet$ \textbf{Scheduling} determines how HE kernels are scheduled for each HE operator (e.g., addition, multiplication, rotation, rescale, bootstrapping). Effective scheduling algorithms aim at reducing kernel invocation counts, saving latency~\cite{demystifying_HE}.

$\bullet$ \textbf{Decomposing} specifies arithmetic and memory operations on individual ciphertexts for HE kernels. For example, NTT algorithms like radix-2 Cooley-Tukey~\cite{kim2024cheddar} and 4-step NTT~\cite{fan2022tensorfhe} are different algorithms in the decomposing layer.

$\bullet$ \textbf{Binding}: Arithmetic and memory operations are translated into hardware-specific programming interfaces (e.g. JAX for TPU), or low-level hardware ISAs (e.g. SIMD ISA for TPU).


Optimizations in packing, mapping and scheduling layers are hardware-agnostic, universally beneficial across hardware platforms. While optimizations in decomposing and binding algorithms are hardware-specific to improve the performance of HE kernels on a specific platform. \compiler primarily addresses inefficiencies in these hardware-specific optimizations for GPU when ported to TPU-like accelerators. We specifically highlight GPU's HE optimizations due to their SoTA performance and hardware similarities with AI accelerators.

\subsection{Architectural Differences: GPU vs. TPU}
\label{sec:arch_diff}

\subsubsection{\textbf{Arithmetic Mismatch - Ratio of MatMul to VecMul Throughput}} While both GPUs and TPUs feature 32-bit integer vector multiplication units (VPUs/CUDA cores) and 8-bit matrix multiplication units (MXUs/Tensor cores), their throughput ratios differ significantly. For instance, one NVIDIA A100’s 8-bit Tensor core offers approximately $4\times$ the throughput of a single 32-bit integer CUDA core (one 32-bit integer MAC is approximated as 16 8-bit MACs), limiting performance gains from vector-to-matrix transformations. Conversely, TPU’s MXUs achieve substantially higher throughput relative to VPUs (e.g., $58\times$ on TPUv4). This creates a massive incentive to reformulate HE algorithms to leverage the MXU, even if the conversion introduces extra computational overhead. 

\subsubsection{\textbf{Memory Manipulation Granularity - Fine-grained vs Coarse-grained}} GPUs feature fine-grained, per-core registers optimized for efficient, element-wise data manipulation. Conversely, TPUs prioritize energy efficiency through a coarse-grained register architecture, consisting of a unified large register file (4 KB) shared across multiple VPUs and MXUs. This design significantly reduces instruction overhead but imposes considerable costs for fine-grained, element-wise data manipulations due to low utilization when mapping small data elements to large registers. Consequently, TPUs inherently favor algorithms that leverage large-tile SIMD computation, while avoiding costly element-wise manipulations.

\subsection{Inefficiency of SoTA Binding Algorithm on TPU}
Current GPU-optimized HE kernels predominantly use 32-bit arithmetic operations~\cite{kim2024cheddar}, including scalar multiplication and modular reduction, which are a natural fit for a GPU's CUDA cores. On a TPU, however, this has dire consequences:

\subsubsection{Idle Matrix Engine for Computing HE Kernels} 
SoTA GPU implementations perform HE kernels such as Basis Conversion (BConv, details in \secref{sec:basis_change_content}) and radix-2 Cooley-Tukey Number Theory Transformation (NTT, details in \secref{sec:ntt}) as sequences of \textit{32-bit scalar multiplications}. 
On TPUs, these 32-bit multiplications are executed on the low-throughput VPU, leaving the high-performance low-precision MXU idle. 


\begin{figure*}[t]
\centering
    \includegraphics[width=\textwidth]{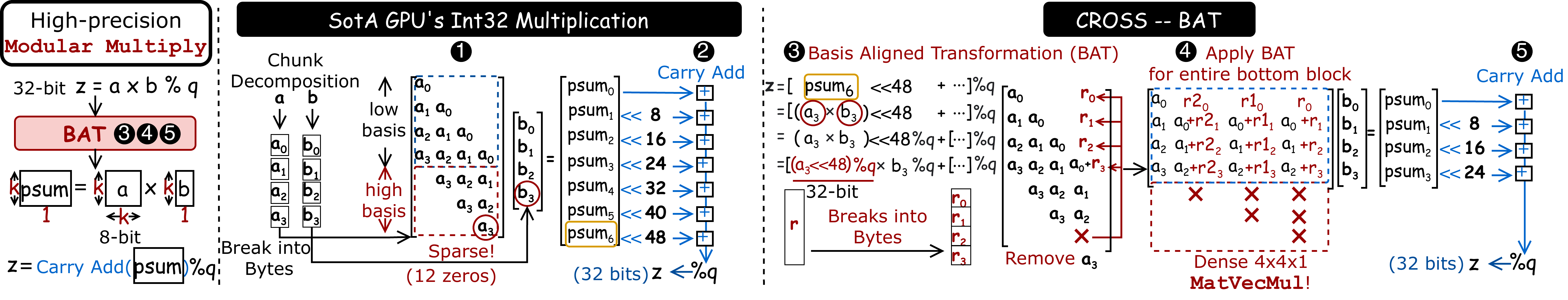}
    \vspace{-7mm}
    \caption{BAT converts high-precision modular scalar multiplication into dense, low-precision matrix multiplication. Compared to the SoTA approach used in GPUs, BAT achieves $2\times$ theoretical computational and memory savings by eliminating redundant zeros of converted matrix. Subscripts indicate chunk indices with each chunk 8-bit long. \RevE{Modular reduction details in} Alg.\ref{alg:montgomery_reduction}.}
    \label{fig:CROSS_Scalar_Flow}
    \vspace{-6mm}
\end{figure*}

\subsubsection{Redundant Zeros in 32-bit Mul.} In GPU's implementation, a high-precision multiplication is decomposed into a sparse matrix-vector multiplication (\texttt{MatVecMul}) of lower-precision chunks (e.g., 8-bit). As shown in \figref{fig:CROSS_Scalar_Flow}\cite{fan2022tensorfhe}, the post-conversion \texttt{MatVecMul} contains a sparse matrix with nearly half zeros, wasting compute and memory resources. 

More efficient method is needed to remove these zeros and convert high-precision arithmetic into dense low-precision matrix operations with no redundant compute and memory. This motivates Basis Aligned Transformation (BAT).

\subsection{Inefficiency of SoTA Decomposing Algorithm on TPU}

HE kernels are rife with data reordering operations that are fundamentally at odds with the TPU's coarse-grained memory system. The reordering required by two latency dominating HE Kernels is analyzed below. Full profiling results see \secref{sec:he_kernels}.

\subsubsection{Number Theoretic Transform (NTT)} It has the highest algorithmic complexity and latency among HE operators.

$\bullet$ {\textit{Radix-2 Cooley-Tukey NTT algorithm}}, optimized for GPUs, relies on fine-grained, bit-complement shuffling at each stage. On a TPU, this requires moving small, non-contiguous data blocks across different memory lanes via the Cross Lane Unit (XLU), resulting in extremely low data tile utilization and prohibitive memory latency. We provide details in \secref{sec:ntt}.

$\bullet$ \textit{4-step NTT algorithm} reformulates the NTT into a chain of matrix multiplications, making it a better candidate for TPUs. However, it introduces both matrix transpose and bit-reverse shuffling. This large-scale data reordering is costly on the TPU architecture, worsening memory bottleneck.

\subsubsection{Automorphism} It requires explicit slot-wise data permutation within a ciphertext, again leading to low VReg utilization and non-trivial memory latency.

These explicit and costly data reordering steps demand a scheduling strategy that removes runtime reordering. It motivates our Memory-Aligned Transformation (MAT).

Post-compiled arithmetic differences among HE workloads, SoTA GPU's HE library and \compiler are listed in \tabref{tab:HE_operator_AI_operator}.

\begin{table}[!htp]\centering
\vspace{-2mm}
\caption{Arithmetic Comparison (GPU's HE lib. vs. CROSS)} \label{tab:HE_operator_AI_operator}
\vspace{-1mm}
\footnotesize
\begin{tabular}{p{5mm}p{24mm}cc}\hline
& \quad \quad \quad Primitive & Precision & Operation \\\hline
 & \quad \quad \texttt{VecModAdd} & $\log_2 q$ & $(\mathbf{a} \+ \mathbf{b}) \text{ mod } q$\\
HE & \quad \quad \texttt{VecModMul} & $\log_2 q$ \circled{1} & $ (\mathbf{a}\times\mathbf{b}) \text{ mod } q$ \\
 & \quad \quad \texttt{ModMatMul} & $\log_2 q$ \circled{2} & $M_{H\times V}\cdot M_{V\times W}$ mod $q$ \\
\hline 
Kernels& \quad \quad \quad \texttt{VecAdd} & (u)int32 & $\mathbf{a} \+ \mathbf{b}$ \\
in & \quad \quad \quad \texttt{VecMul} & (u)int32 & $\mathbf{a} \times \mathbf{b}$   \\
GPU's & \quad \ Sparse \texttt{MatMul} & (u)int8 & $M_{(H+K-1)\times V}\cdot M_{V\times W}$  \\
HE & \quad \quad \texttt{Transpose} & 32 bit & $M_{H\times V} \rightarrow M_{V\times H}$ \\
Library& \quad \ \texttt{Permutation} & 32 bit & Across Lane (\figref{fig:tpu_arch_new}) \\
\hline
& \quad \quad \quad \texttt{VecAdd} & (u)int32 & $\mathbf{a} \+ \mathbf{b}$ \\
CROSS & \quad \quad \quad \texttt{VecMul} & (u)int32 & $\mathbf{a} \times \mathbf{b}$   \\
  & \quad \ Dense \texttt{MatMul} & (u)int8 & $M_{H\times V}\cdot M_{V\times W}$   \\
\hline
\end{tabular}

\circled{1}: $28\leq \log{q}\leq 59$ under 128-bit security level; Intermediate results need $2\log q$ bits, far exceeding precision range of GPUs or AI accelerators.\\
\circled{2}: intermediate results need $2\log{q}+\log{V}$ bits to avoid precision overflow. \\
\vspace{-4.5mm}
\end{table}

\section{CROSS Methods}
This section introduces two key contributions, BAT and MAT, in resolving arithmetic mismatch and memory manipulation granularity challenge categorized in \secref{sec:arch_diff}.

\subsection{Basis Aligned Transformation (BAT)}
\label{sec:BAT}

BAT, our innovation in binding algorithm, is designed to efficiently transform high-precision modular integer arithmetic into dense low-precision matrix multiplication, enabling the TPU's MXU, with $\sim O(100)$ times higher throughput than its VPU, to effectively accelerate primitives in \tabref{tab:HE_operator_AI_operator}. 

The key idea of BAT lies in offline pre-computing known parameters—such as twiddle factors in NTT, primes in BConv, and evaluation keys in relinearization—to minimize runtime overhead in transformed low-precision matrix multiplication. We first illustrate BAT's methodology using a concrete example of transforming 32-bit integer multiplications into efficient 8-bit \texttt{MatVecMul}. We then highlight BAT's advantages over the SoTA 32-bit multiplication approach used in GPUs and explain its applicability to enhancing Montgomery reduction and high-precision Modular matrix multiplication (\texttt{MatModMul}).

\subsubsection{BAT Methodology and Illustration}
\label{sec:bat}

In the SoTA high-precision scalar multiplication used by GPUs\cite{fan2022tensorfhe, tensorfhe_plus}, two inputs of 32-bit standard modular multiplication, $a$ and $b$, are first broken into four 8-bit chunks $(a_0, a_1, a_2, a_3)$ and $(b_0, b_1, b_2, b_3)$. Then it performs all-to-all chunk multiplications to produce products (termed as partial sum, noted as psum) that contribute to different ``output bases" (i.e., powers of $2^8$). For instance, $a_3 {\times} b_3$ contributes to the final sum with a basis of $2^{48}$. The SoTA approach maps the computation from input chunks to partial sums as a sparse \texttt{MatVecMul}, as shown in \figref{fig:CROSS_Scalar_Flow} (\ding{182}), where left matrix is a toeplitz matrix of all chunks of $a$. It contains $12/(4\times7)\approx43\%$ zeros and leads to redundant compute and memory overhead. Further, it explicitly computes all seven partial sums and accumulates them via a long chain of carry-and-add operation (\ding{183}).

BAT recognizes a critical opportunity: since the final result must be taken modulo $q$, the contributions of high-basis terms can be calculated and ``folded" into the low-basis terms in compile time, i.e. BAT applies modulo $q$ to the shifted sparse left matrix, which converts it into a smaller, dense matrix that directly computes the final low-basis coefficients.

Specifically, elements in each left matrix row (\ding{182}) contribute to distinct bit ranges in the final 64-bit partial sum ($psum$). If we use $bp$ as bit precision of matrix multiplication, which is 8 for TPU, then each row $k \in [0, 6]$ has an associated bit range from $8k$ to $8(k + 2) \cdot bp + 2$, with an output basis of $2^{k \cdot bp}$. Post modular reduction, the final result $z$ (\ding{183}) retains only the four lowest bases (lower 32 bits, for $k \in [0,3]$), rendering higher-basis contributions intermediate. BAT directly applies modular reduction to elements in sparse left matrix contributing to these higher-basis partial sums, realigning them to the lowest bases. For instance, in the case of $a_3 \times b_3 = psum^6$, BAT transforms $psum^6$ by computing $r = (a_3 << 48) \mod q$ and subsequently decomposing it into four 8-bit chunks ($r_0, r_1, r_2, r_3$). These chunks are then added back to the top blue dash box (\ding{184}).

By systematically applying BAT to all elements in the block of high basis (red dash box), the sparse left matrix shrinks into a smaller dense matrix (\ding{182}$\rightarrow$\ding{185}), achieving a $\sim2\times$ theoretical computational and memory saving in the \texttt{MatMul}. Further, the length of carry-add chain is reduced from seven to four (\ding{183}$\rightarrow$\ding{186}), \RevE{saving latency when being mapped to VPU.} 

Overall, BAT adds an offline pre-computation overhead to eliminate runtime redundancy  and reduce the size of temporal reduction, enhancing computation and memory efficiency. The above procedures are illustrated in \figref{fig:CROSS_Scalar_Flow} with detailed explanation detailed in Alg. \ref{alg:HP_on_LP_special_new} in the appendix.

\subsubsection{Math of BAT}
For arbitrary precision input value $a,b$, each with $K$ bytes. Assuming the value of $a$ is preknown (e.g. twiddle parameters, evaluation key, parameters in basis switch), modular multiplication of $a$ and $b$ is reformed as: 
\label{sec:accelerated_bat}
\begin{align}
& a\times b\bmod q = \left( \sum_{i=0}^{i=K-1} \underbrace{a}_{\text{K bytes}} \times (\underbrace{b_{i}}_{\text{one byte}}\times 2^{i})\right) \bmod q  \label{eq:line2}\\ 
& = \left( \sum_{i=0}^{i=K-1} \underbrace{\left( a \times 2^{i} \bmod q \right)}_{\text{calculated offline as }a_{i} \text{ ($K$ bytes)}} \times \underbrace{b_{i}}_{\text{one byte}} \right) \bmod q \\
& = \left( \sum_{i=0}^{i=K-1} \underbrace{a_i}_{\text{ $K$ bytes}} \times \underbrace{b_{i}}_{\text{one byte}} \right) \bmod q \label{eq:line6} \\
& = \left( \sum_{i=0}^{i=K-1} \underbrace{(\sum_{j=0}^{j=K-1} a_{j,i}\times 2^{8j})}_{\text{decomposes $a_i$ as bytes}} \times \underbrace{b_{i}}_{\text{one byte}} \right) \bmod q \label{eq:line8} \\
& = \left( \sum_{j=0}^{j=K-1} \underbrace{( \sum_{i=0}^{i=K-1} a_{j,i} \times b_{i})}_{\text{8-bit matrix multiplication}}\times 2^{8j} \right) \bmod q \label{eq:line9} \\ 
& = \sum \left( \underbrace{\begin{bmatrix}
a_{0,0}  & \cdots & a_{0,K-1} \\
a_{1,0}  & \cdots & a_{1,K-1} \\
\vdots        & \ddots & \vdots        \\
a_{K-1,0} & \cdots & a_{K-1,K-1}
\end{bmatrix}}_{ K\times K \text{ 8-bit matrix}} \times \underbrace{\begin{bmatrix}
b_{0} \\
b_{1} \\
\vdots \\
b_{K-1}
\end{bmatrix}}_{K\times 1 \text{ 8-bit}} \times \begin{bmatrix}
2^0 \\
2^8  \\
\vdots \\
\vdots 
\end{bmatrix} \right) \bmod q \\
& = \left( \sum_{j=0}^{j=K-1} ( \underbrace{psum_j}_{16+\log_2(K)\ \text{bits}} \times 2^{8j} ) \right) \bmod q
\end{align}
BAT offline applies modular reduction to preknown $K$-byte parameter $a$ in Eq. (\ref{eq:line2}), converting it into Eq. (\ref{eq:line6}). BAT then schedules post converted computation as a low-precision matrix multiplication with a carry propagation. This conversion takes $O(N)$ time for converting a matrix with $O(N)$ elements. 

\subsubsection{Accelerating Core HE Kernels with BAT}
\label{q:re_c1}
\paragraph{BAT Lazy Modular Reduction for Scalar} BAT can be applied to modular reduction by transforming it into a matrix multiplication (MatMul) followed by carry propagation (\secref{sec:bat_lazy}). However, the resulting MatMul has reduction dimension $K$, which favors small matrix engine like the tensor cores in GPUs but underutilizes the MXU. Therefore, we adopt optimized Montgomery reduction (Alg.~\ref{alg:montgomery_reduction}) and Barrett reduction (Alg.~\ref{alg:barrett}) and map both to the VPU.

\begin{algorithm}
\caption{Optimized Montgomery Reduction (64${\rightarrow}$32 bits)} 
\label{alg:montgomery_reduction} 
\begin{algorithmic}[1] 
\Require $z {\in} [0, 2^{64})$, $q{<}2^{32}$, $q_{lo} {=} q \pmod{2^{16}}$, $q_{hi} {=} \lfloor q / 2^{16} \rfloor$
\Ensure $B \equiv z \cdot 2^{-32} \pmod{q}$, $B \in [0, 2q)$ 
\State $z_{lo} \leftarrow z \pmod{2^{32}}$, $z_{hi} \leftarrow \lfloor z / 2^{32} \rfloor$ \hfill $\triangleright$ Split 64-bit input
\State $t \leftarrow (z_{lo} \cdot q^{-1}) \pmod{2^{32}}$ \hfill $\triangleright$ Low 32-bit product
\State $t_{lo} \leftarrow t \pmod{2^{16}}$, $t_{hi} \leftarrow \lfloor t / 2^{16} \rfloor$ \hfill $\triangleright$ Split $t$ for 16-bit mults
\Statex
\noindent \hspace{-6mm} \textbf{Compute upper 32 bits of $(t \cdot q)$ using 16-bit primitives:}
\State $p_{hi} \leftarrow t_{hi} \cdot q_{hi}$; $p_{lo} \leftarrow t_{lo} \cdot q_{lo}$; $p_{m,hi} \leftarrow t_{hi} \cdot q_{lo}$; $p_{m,lo} \leftarrow t_{lo} \cdot q_{hi}$
\State $mid_{lo} \leftarrow p_{m,hi} + p_{m,lo} + \lfloor p_{lo} / 2^{16} \rfloor$
\State $mid_{hi} \leftarrow \lfloor p_{m,hi} / 2^{16} \rfloor + \lfloor p_{m,lo} / 2^{16} \rfloor + \lfloor mid_{lo} / 2^{16} \rfloor$
\State $t_{final} \leftarrow p_{hi} + mid_{hi}$ \hfill $\triangleright$ This equals $\lfloor (t \cdot q) / 2^{32} \rfloor$
\State $B \leftarrow z_{hi} + q - t_{final}$ \hfill $\triangleright$ Result in $[0, 2q)$
\State \Return $B \pmod{2^{32}}$
\end{algorithmic}
\end{algorithm}



\begin{figure}[h]
    \centering
    \vspace{-6mm}
    {\includegraphics[width=\columnwidth]{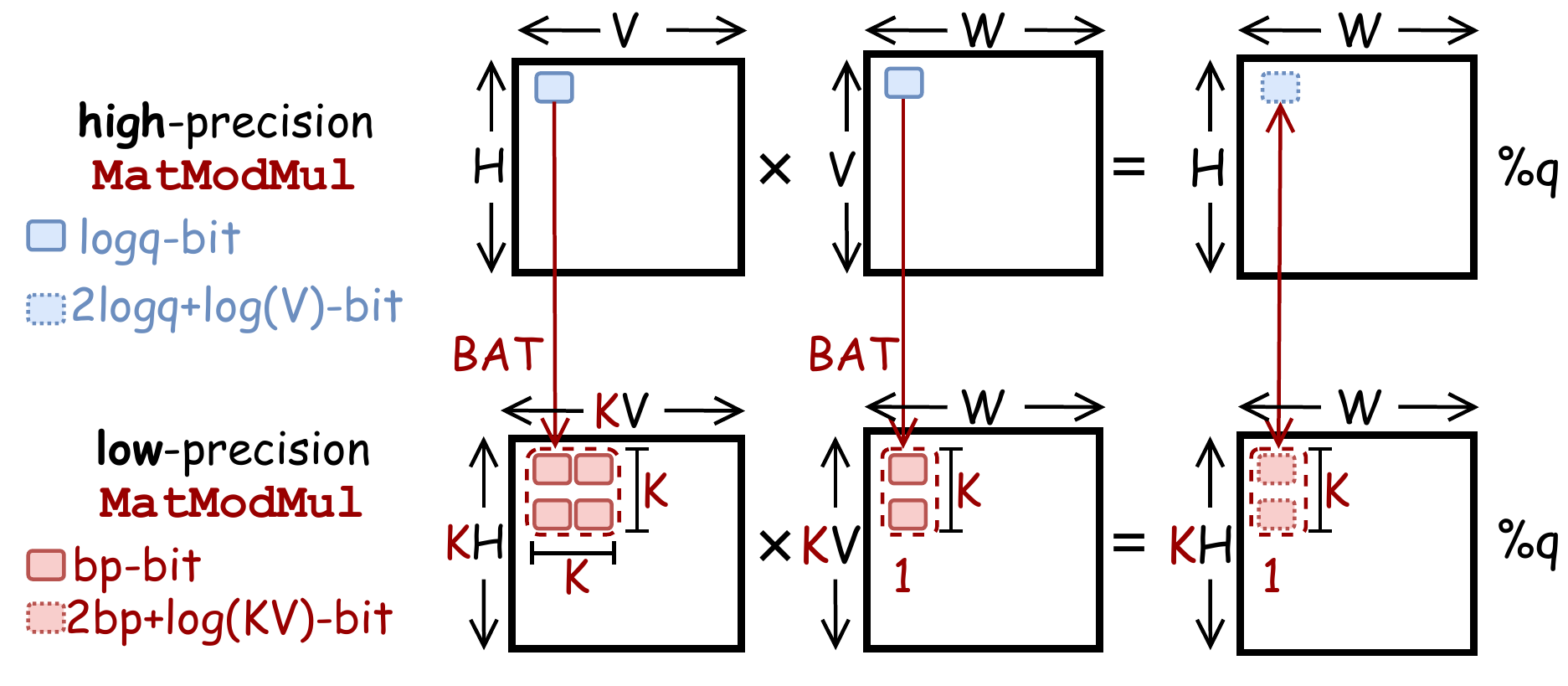}}
    \caption{BAT could be applied to convert each individual elements of high-precision \texttt{MatModMul} to transform it into low-precision \texttt{MatMul}. This enlarges dimensions by a factor of $K = \lceil \frac{\log_2 q}{bp} \rceil$, enhancing MXU utilization. Specifically, \compiler employs BAT to represent each high-precision ($\log_2 q$-bit) element from the input matrices as either a $K \times K$ array or a $K$-length vector composed of lower-precision ($bp$-bit) elements. The resulting element has $2bp + \log_2(KV)$ bits, accounting for precision expansion during reduction (Alg. \ref{alg:bat_deployment}).}
    \label{fig:ModMatMul}
    \vspace{-2mm}
\end{figure}

\paragraph{High-Precision \texttt{MatModMul} in BConv} BConv requires \texttt{VecModMul} and \texttt{MatModMul} in two steps (details in \secref{sec:basis_change_content}).
32-bit \texttt{ModMatMul} in step 2 dominates latency in BConv. Abstractly, this operation can be represented as an 32-bit ($H, V, W$)-\texttt{ModMatMul}, which could only utilize the low-throughput VPU, failing to exploit inherent matrix multiplication data reuse, thus limiting performance. BAT addresses this limitation by reformulating the operation into an 8-bit ($KH, KV, W$)-\texttt{ModMatMul} (Alg. \ref{alg:bat_deployment}), suitable for MXU acceleration. This transformation (1) improves the throughput and (2) leverages the inherent data reuse capability of MXU to reduce memory traffic, mitigating memory boundness. 

\begin{algorithm}
\caption{Applying BAT to High-Precision ModMatMul}\label{alg:bat_deployment} 
\begin{algorithmic}[1]
\Require Modulus $q$; Preknown matrix $A \in \mathbb{Z}_{q}^{H \times V}$, input data matrix $B \in \mathbb{Z}_{q}^{V \times W}$; $bp$: MXU bit precision; $K \leftarrow \lceil \frac{\log_2 q}{bp}\rceil$.
\Ensure $Z \in \mathbb{Z}_{q^2\times V}^{H \times W}$ (high-precision MatMul result).

\item[]
\item[]
\hspace{-2em} \textbf{\textsc{ChunkDecompose}}{($a$)}{}$\rightarrow \left[a_{k}\right]_{0\le k < K}$
\State \textbf{for} $k = 0$ \textbf{to} $K-1$:
\State \quad $a_k \leftarrow (a \gg (k \cdot bp)) \ \& \ (2^{bp} - 1)$ \hfill $\triangleright$ Mask $k$-th $bp$ bits.
\State \textbf{Return} $\left[a_{k}\right]_{0\le k < K}$

\item[]
\item[]
\hspace{-2em} \textbf{\textsc{ChunkMerge}}{($\left[a_{k}\right]_{0\le k < K}$)}{} $\rightarrow a$
\State $a \leftarrow 0$
\State \textbf{for} $k = 0$ \textbf{to} $K-1$:
\State \quad $a \leftarrow a + (a_k \ll (k \cdot bp))$
\State \textbf{Return} $a$
\item[]
\item[]
\hspace{-2em} \textbf{\textsc{DirectScalarBAT}}{($a$)}{} $\rightarrow M_{K \times K}$
\State Initialize $M \leftarrow \mathbf{0}^{K \times K}$
\State \textbf{for} $j = 0$ \textbf{to} $K-1$: \hfill $\triangleright$ Iterate columns (input basis $2^{j \cdot bp}$)
\State \quad  $val \leftarrow (a \ll (j \cdot bp)) \pmod q$ \hfill $\triangleright$ Shift and reduce
\State \quad  $[chunk_0, \dots, chunk_{K-1}] \leftarrow$ \textsc{ChunkDecompose}($val$)
\State \quad  \textbf{for} $i = 0$ \textbf{to} $K-1$: \hfill $\triangleright$ Fill rows (output basis)
\State \quad \quad $M[i, j] \leftarrow chunk_i$ \hfill $\triangleright$ Assign $i$-th chunk to row $i$
\State \textbf{Return} $M$

\item[]
\item[]
\hspace{-2em} \textbf{\textsc{OfflineCompileLeft}}{($A_{H \times V}$)}{} $\rightarrow A_{dense}$  \hfill $\triangleright$ Offline
\State Initialize $A_{dense} \leftarrow \mathbf{0}^{KH \times KV}$
\State \textbf{for} $h = 0$ \textbf{to} $H-1$:
\State \quad \textbf{for} $v = 0$ \textbf{to} $V-1$:
\State \quad \quad $M_{sub} \leftarrow$ \textsc{DirectScalarBAT}($A[h, v]$)
\State \quad \quad $A_{dense}[hK : (h+1)K, \ vK : (v+1)K] \leftarrow M_{sub}$ 
\State \hfill $\triangleright$ Embed $K \times K$ block
\State \textbf{Return} $A_{dense}$

\item[]
\item[]
\hspace{-2em} \textbf{\textsc{RuntimeCompileRight}}{($B_{V \times W}$)}{} $\rightarrow B_{dense}$
\State Transforms Right Matrix to $KV \times W$ layout \figref{fig:ModMatMul}.
\State Initialize $B_{dense} \leftarrow \mathbf{0}^{KV \times W}$
\State \textbf{for} $v = 0$ \textbf{to} $V-1$:
\State \quad \textbf{for} $w = 0$ \textbf{to} $W-1$:
\State \quad \quad $[b_0, \dots, b_{K-1}] \leftarrow$ \textsc{ChunkDecompose}($B[v, w]$)
\State \quad \quad \textbf{for} $k = 0$ \textbf{to} $K-1$:
\State \quad \quad \quad $B_{dense}[vK + k, w] \leftarrow b_k$ \hfill $\triangleright$ Stack chunks vertically
\State \textbf{Return} $B_{dense}$
\item[]
\item[]
\hspace{-2em} \textbf{\textsc{Main-FullMatMul}}{($A, B$)}{} $\rightarrow Z$
\State $A_{dense} \leftarrow$ \textsc{OfflineCompileLeft}($A$) \hfill $\triangleright$ $KH \times KV$
\State $B_{dense} \leftarrow$ \textsc{RuntimeCompileRight}($B$) \hfill $\triangleright$ $KV \times W$
\State $Z_{chunk} \leftarrow A_{dense} @ B_{dense}$ \hfill $\triangleright$ Low-precision MatMul (MXU)
\State Initialize $Z \leftarrow \mathbf{0}^{H \times W}$
\State \textbf{for} $h = 0$ \textbf{to} $H-1$:
\State \quad \textbf{for} $w = 0$ \textbf{to} $W-1$:
\State \quad \quad $Z[h,w] \leftarrow$ \textsc{ChunkMerge}($[\, Z_{chunk}[hK,w],\dots,$ $Z_{chunk}[(h{+}1)K{-}1),w] \,]$)
\State \textbf{Return} $Z$
\end{algorithmic}
\end{algorithm}


\subsection{Memory Aligned Transformation (MAT)}
\label{sec:MAT}

MAT is a scheduling optimization that removes explicit runtime memory reordering overhead by \textit{integrating data transformation directly into computation}. Its core principle is to permute parameter offline in compile time, thereby ensuring runtime computations inherently produce outputs in the desired data layouts. This strategy effectively eliminates runtime memory reordering costs.

\subsubsection{MAT Key Idea and Illustration}

MAT leverages the insight that any reordering operation on a one-dimensional vector can be represented as multiplication with a ``permutation matrix", which refers to a matrix containing exactly one non-zero element per row and per column. By applying the permutation matrix offline to pre-known parameters (e.g., twiddle factors), the post MAT kernel generates the output in the expected order without incurring runtime reordering. We demonstrate MAT's efficacy using two representative HE workloads: \texttt{VecMul}, which requires element permutation, and \texttt{MatMul}, which necessitates matrix transposition (\figref{fig:MAT_illustration}). MAT is also effective for multi-dimensional tensors by creating one permutation matrix per dimension offline. We fuse all permutation matrices into one.

\subsubsection{Applying MAT to 4-step NTT for Layout Invariance}
\label{sec:MAT_NTT}
\paragraph{Transpose Elimination}
The SoTA GPU's tensor core centric 4-step NTT~\cite{warpdrive} consists of reforming $N$-input data into $(R,C)$ matrix and performing four steps:
\squishlist
\item Perform \(R\)-input NTTs on each of the \(C\) columns.
\item {Transpose resulting }\(R\times C\) {matrix to obtain a} \(C\times R\) {layout}.
\item Multiply by precomputed twiddle factors.
\item Perform \(C\)-input NTTs on each of the \(R\) rows.
\squishend
\insertFigure{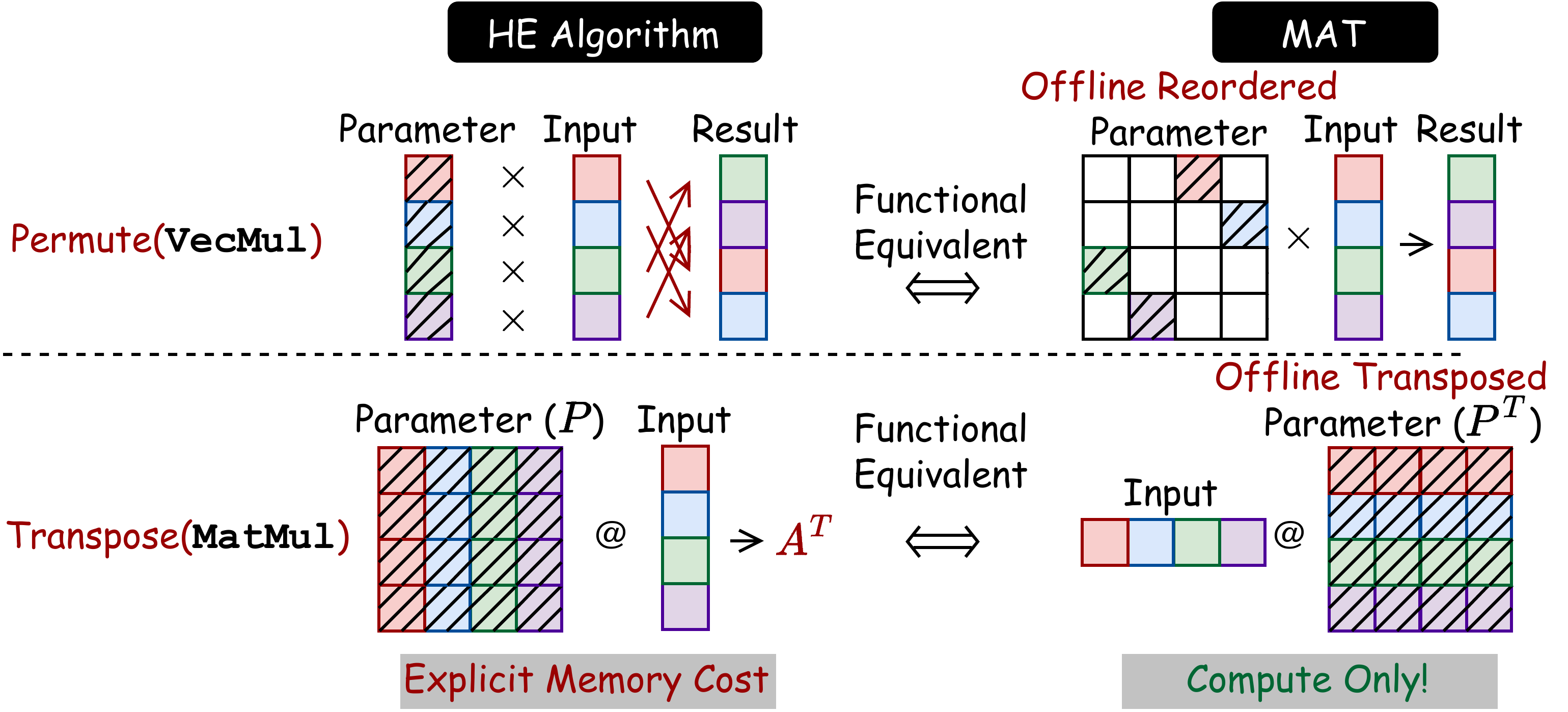}{MAT Illustration for \texttt{Permute(VecMul)} and \texttt{Transpose(MatMul)}. MAT moves explicit memory reordering to compiler time by applying reordering directly on preknown parameters offline for runtime latency saving.}

\insertWideFigureRatio{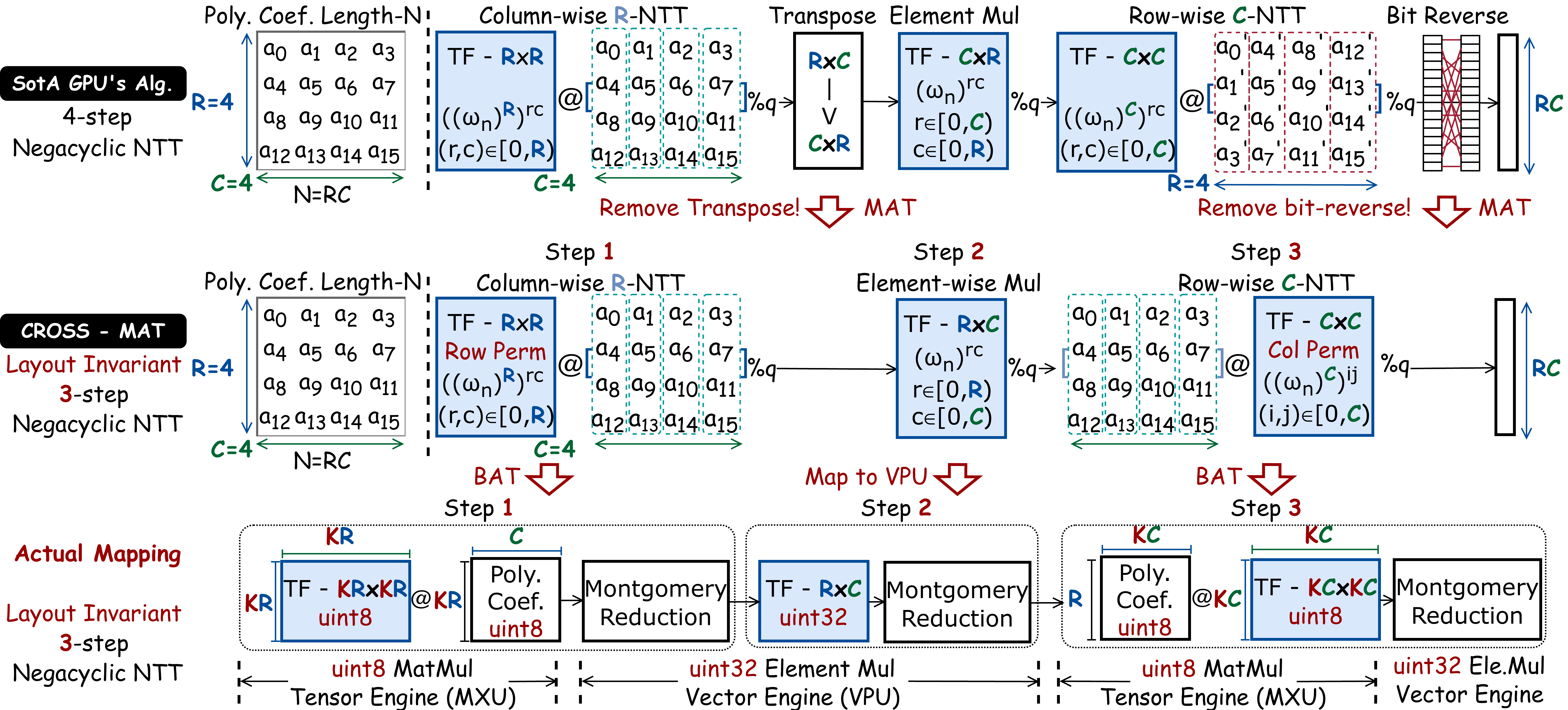}{\compiler \ converts \textit{high-precision NTT} into a combination of low-precision matrix multiplications and element-wise multiplications to fully exploit the MXU and VPU. Row 1 shows the conventional 4-step NTT algorithm~\cite{fan2022tensorfhe}; Row 2 illustrates how the Memory Aligned Transformation (MAT) eliminates the explicit transpose and bit-reverse permutation to keep data layout invariant in NTT. Row 3 details the mapping of the MAT-optimized algorithm onto the TPU architecture. Here, TF denotes twiddle factor. Twiddle factors in step 1 and 3 become the same when $R=C$, such that only one TF matrix is needed.}{0.9}
4-step NTT transposes input data from row-major layout into column-major layout, incurring explicit memory overhead. To eliminate it, we apply \textit{MAT} to embed the transpose into the second matrix multiplication as shown in \figref{fig:MAT_illustration}. Leveraging the identity $(A@B)^T=B^T@A^T$, MAT frees the demand of column-major layout back to its input row-major layout, effectively removing transpose. Further, MAT exploits the symmetry structure of twiddle factors (TF), i.e. $(TF^{C}_{C\times C})^T = TF^{C}_{C\times C}$). Therefore, MAT merely interchanges the multiplication order of coefficients and twiddle factors of the second matrix multiplication.

\paragraph{Bit-reverse Shuffling Elimination}
Typical negacyclic NTT produces results in the bit-reverse order. We further apply MAT to embed the bit-reverse reordering into offline coefficients permutation, i.e., permute the rows of Step 1 and Step 2 matrices, and the columns of the Step 3 matrix, using bit-reversal indices. This ensures runtime computation to directly generate the results in the bit-reversed order, i.e., purely using two matrix multiplications to generate a bit-reverse reordered output. Such offline permutation works for bit-reverse order with power of 2 elements, and being applicable to all NTTs and INTTs used in HE. This step ensures ``layout-invariant" throughout the entire NTT/INTT process, thus we call it \textit{layout invariant 3-step negacyclic NTT}, as shown in \figref{fig:NTT_overall_flow}. It's detailed formula is:
\begin{equation*} 
( \underbrace{(P_{\pi(R)} @ TF^{R}_{R\times R})}_{\text{offline row permutation}} @ a_{R\times C} ) \cdot \underbrace{(P_{\pi(R)} @ TF_{R\times C})}_{\text{offline row perm.}}\big) @ \underbrace{(TF^{C}_{C\times C} @ P_{\pi(C)})}_{\text{offline column perm.}}
\end{equation*}

Here, \( @ \) denotes matrix multiplication and \(\cdot\) represents element-wise multiplication. $TF^{k}_{R\times C}$ means a twiddle factor matrix of $[((\omega_n)^k)^{rc}], r\in[0,R), c\in[0,C)$. $\omega_n$: primitive $n$-th root of unity.
$P_{\pi(R)} = \Big[ \delta_{c, \rho(r)} \Big]_{r,c \in [0, R)}$ and $P_{\pi(C)} = \Big[ \delta_{r, \rho'(c)} \Big]_{r,c \in [0, C)}$ denote bit-reversal permutation matrix. $\rho(i)$ and $\rho'(i)$ denote the bit-reversal of index $i$ for size $R$ and $C$, separately. $\delta_{i,j}$ is the Kronecker delta ($1$ if $i=j$, $0$ otherwise).

\paragraph{Mapping to TPU}
In layout-invariant 3-step negacyclic NTT, high-precision matrix multiplications in steps 1 and 3 are transformed into low-precision matrix multiplications by BAT, which will be executed by MXU to achieve high throughput. Other operators are mapped to VPU (\figref{fig:NTT_overall_flow}).

Integrating MAT and BAT, CROSS efficiently exploits both MXU and VPU in sequential order listed from left to right in the third row of \figref{fig:NTT_overall_flow}. The resulting algorithm achieves an effective computational complexity of $O(N\sqrt{N})$ while completely avoiding costly data reordering. Although this complexity is higher than the radix-2 Cooley-Tukey NTT ($O(N \log N)$), the dramatic throughput advantage of MXU results in superior overall NTT throughput and energy efficiency, making TPUs the SoTA throughput engine for NTT.

\section{Evaluation}
\label{sec:evaluation}
We investigate the performance benefits of BAT and MAT exclusively for latency dominating HE operators. We also evaluate the throughput per watt (energy efficiency) of CROSS framework on Google's TPU architectures against SoTA implementations on various platforms including CPUs, GPUs, FPGAs, and HE ASICs. We analyze the reasons behind performance gap to the dedicated HE ASICs, and summarize future directions to improve AI ASIC's HE performance.

\subsection{Methodology}
We select Google's TPU as a representative ASIC AI accelerator and enable deployment on real TPUs by converting HE operators into \texttt{MatMul} and \texttt{VecMul} and leveraging JAX~\cite{jax2018github} to program TPUs. We report the latency obtained from trace viewer in XLA profiler~\cite{XLA_Profiler}. For bootstrapping, MNIST inference and Logistic Regression, the estimated latency is obtained by multiplying the overall number of HE kernels invocations with each profiled realistic latency, which represents the worst case latency as it assumes no pipeline or fusion. 

\textbf{Security Parameter Selection:} Security parameter setup has significant effect on the performance and accuracy. Given TPU's micro-architecture is mainly designed for optimizing performance of low-precision integer (up-to 32 bits) and it implements 32-bit registers, we choose the security parameter with $\log_2q<32$ for better performance. For security parameters requiring moduli precision exceeding 32 bits, we employ double rescaling\cite{cheon2019full} to discard two sub-moduli ($\log_2q < 32$) per level, doubling the number of constituent moduli.

\textbf{CROSS Configuration:}\label{q:rc_c3}
Unless otherwise stated, all experiments use a 128-bit security configuration with $\log_2 q = 28$, $L = 51$, $D_{\text{num}} = 3$, and $N = 2^{16}$ (Set~D in \tabref{tab:ntt_throughput_setup}). Each 28-bit coefficient is stored in a 32-bit integer, and decomposed into four 8-bit chunks to match MXU operand precision when aiming at using MXU. CROSS employs a layout-invariant 3-step NTT, with $(R,C)\in\{(128,512),(256,256),(512,128)\}$ selected to efficiently map onto MXU. For standalone NTT evaluation, we fix $(R,C)=(128,\lfloor N/128\rfloor)$, explicitly setting one dimension to 128 (the number of lanes) to ensure full VReg utilization at small problem sizes, such as $\sqrt{N}<128$. For HE operator evaluation, we sweep three $(R,C)$ configurations and report results using the best-performing one.


\textbf{Workload:} We adopt ML workloads without bootstrapping including MNIST and Logistic Regression\footnote{HELR~\cite{helr_han_2019} is a binary classification model using logistic regression. We trained the model for 32 iterations, each with a batch containing 1024 14$\times$14-pixel MNIST images, where an iteration is a gradient update step with a single batch, and report average execution time per iteration.}~\cite{helr_han_2019}, and four backbone HE operators as workloads, including HE-Add, HE-Mult, Rescale, and Rotate~\cite{CKKS}. CROSS's HE operator implementations are verified against OpenFHE's leveled ckksrns~\cite{openfhe}, achieving the same accuracy and generality as the OpenFHE's implementation. We adopt the packed bootstrapping algorithm defined in~\cite{agrawal2023mad}.

\textbf{Baselines:} \textit{For TPU}, the baseline refers to using the SoTA GPU's decomposing and binding algorithm, including (1) breaking high-precision modular scalar multiplication as low-precision modular \texttt{MatMul} illustrated in \figref{fig:CROSS_Scalar_Flow} and (2) 4-step NTT. \textit{For other platforms }like CPU~\cite{mouchet2020lattigo}, GPU~\cite{fan2022tensorfhe,kim2024cheddar}, FPGA~\cite{agrawal2022fab}, and HE ASIC\cite{CraterLake,basalisc_chip}, the security configuration leading to the best performance for each platform is chosen. We further choose the latest works which report NTT throughput and latency of HE kernels in each evaluation, respectively.

\textbf{AI-ASIC Devices Setup and Metrics}. We deploy CROSS on a single-host TPU virtual machine (TPU-VM) of different generations, including TPUv4, TPUv5e, TPUv5p, and TPUv6e. We use TPUv6e by default. The detailed number of tensor core and performance specifications per Tensor Core (TC) are listed in \tabref{tab:ntt_throughput_setup}. A TPU-VM refers to a group of TPU chips that share the same CPU host, e.g. four chips being arranged as 2{$\times$}2 torus to be controlled by a single CPU.

\label{q:rc_c2}\textbf{Metric:} For fair comparison with prior works\cite{kim2024cheddar,agullódomingo2025fideslibfullyfledgedopensourcefhe,basalisc_chip,CraterLake,warpdrive}, we report latency and throughput per watt following methodology below. 
For \textit{latency} of each HE kernel, we run the same kernel on each tensor core and report amortized single-batch latency. For \textit{throughput per watt} (\textit{energy efficiency}), we measure the number of kernels completed per second under a TC configuration that matches the thermal design power (TDP) of the comparison device. Specifically, we constrain the system to 4 TCs when comparing against the Alveo U280 (FPGA), A100 (GPU), and BASALISC/CraterLake (ASIC) architectures, and scale to 2 TCs for the AMD 9950X3D (CPU), and 8 TCs for RTX 4090 (GPU) baselines and HEAP\cite{heap}, respectively. We measure device energy efficiency using the average latency of dominating HE-Mult and Rotate.

\begin{table}[!htp]\centering
\caption{TPU setup and NTT evaluation setup.}\label{tab:ntt_throughput_setup}
\vspace{-2mm}
\resizebox{\columnwidth}{!}{ 
\begin{tabular}{cccccc}\hline
Hardware &\textbf{TPUv4} &\textbf{TPUv5e} &\textbf{TPUv5p} &\textbf{TPUv6e} \\\hline
\# JAX Logical Devices & 4 & 4 & 4 & 8 \\
Setup &v4-8 &v5litepod-4  &v5p-8 &v6e-8\\
\# Tensor Cores & 8 & 4 & 8 & 8 \\ \hline
GFLOPs &139800 &202700  &236700 &918000 \\
HBM BW (GiB/s) &572 &763  &1287 &1526 \\
VMEM Read BW (GiB/s) &2003 &17166 &20027 &21696 \\
VMEM Write BW (GiB/s) &1001 &5722 &6676  &15020 \\\hline\hline
Configuration & \textbf{$\log_2Q$} & Degree & $\log_2q$ & \#Limbs \\\hline
\textbf{Set A} & 109 & $2^{12}$ & 28 & 4 \\
\textbf{Set B} & 218 & $2^{13}$ & 28 & 8 \\
\textbf{Set C} & 438 & $2^{14}$ & 28 & 15 \\
\textbf{Set D} & 1904 & $2^{16}$ & 28 & 51 \\
\hline
\end{tabular}}
TPUv6e (Set D) is the default configuration used by CROSS in evaluations. \\
Specifications (FLOPs and BW) obtained from XProf for \textbf{one tensor core}.\\ 
TPUv6e offers higher GFLOPs as it has 256$\times$256 systolic array.
\vspace{-4mm}
\end{table}

\subsection{Evaluating Performance of Individual CROSS Optimization}

\subsubsection{BAT Evaluation} \label{q:rf_q4}
\paragraph{High-precision \texttt{MatModMul}  --  BAT vs. baseline}
\texttt{MatModMul} in SoTA GPU's HE algorithm mainly arises from NTT and INTT. NTT-based \texttt{MatModMul} typically involves square matrices of size $2^{8}$ to $2^{11}$. On a single TPUv6e Tensor Core, BAT achieves a speedup of up to 1.62$\times$ over the low-precision sparse matrix multiplication baseline. This efficiency gain is attributed to the elimination of redundant computations on zero elements and the removal of data-type conversion overhead for static input parameters (e.g., twiddle factors). The lower speedup observed for smaller matrix dimensions is due to their memory-bound nature. In this regime, performance for both BAT and the baseline is constrained by HBM bandwidth rather than peak computational throughput. 

\begin{table}[!h]\centering
\vspace{-2mm}
\caption{BAT vs. baseline on $M_{H\times V}@ M_{V\times W} \mod q$}\label{tab:bat_perform}
\vspace{-1mm}
\scriptsize
\begin{tabular}{cccccc}\hline
H &V &W &Baseline &BAT &speedup \\\hline
512 &256 &256 &6.00 $\mu s$ &4.57 $\mu s$ & \textbf{1.31$\times$} \\
1024 &256 &256 &9.40 $\mu s$ &6.88 $\mu s$ & \textbf{1.37$\times$} \\
2048 &256 &256 &15.43 $\mu s$ &11.06 $\mu s$ & \textbf{1.39$\times$} \\
4096 &256 &256 &29.09 $\mu s$ &20.14 $\mu s$ & \textbf{1.44$\times$} \\
1024 &512 &512 &20.58 $\mu s$ &16.32 $\mu s$ & \textbf{1.26$\times$} \\
2048 &512 &512 &38.49 $\mu s$ &28.48 $\mu s$ & \textbf{1.35$\times$} \\
1024 &1024 &1024 &59.13 $\mu s$ &40.69 $\mu s$ & \textbf{1.45$\times$} \\
2048 &1024 &1024 &113.91 $\mu s$ &81.71 $\mu s$ & \textbf{1.39$\times$} \\
2048 &2048 &2048 &365.28 $\mu s$ &224.80 $\mu s$ & \textbf{1.62$\times$} \\\hline
\end{tabular}
\vspace{-2mm}
\end{table}

\paragraph{\texttt{BConv}  --  BAT vs. TPU baseline}
Leveraging BAT, the step 2 of BConv is converted from high-precision \texttt{VecModMul} into low-precision skewed \texttt{MatModMul}, with one dimension equal to degree $N$ and the other being the number of limbs $KL$ or $KL'$ in \tabref{tab:bc_skew_shape}, resulting in ($N$, $KL$, $KL^\prime$)-\texttt{MatModMul}. When the reduction dimension in MatMul is not divisible by 128, CROSS pads zeros, leading to partial MXU utilization. Such conversion of using high-throughput MXU brings \textbf{$\leq7.16\times$} speedup on one tensor core (\tabref{tab:bc_skew_shape}).

\begin{table}[!h]\centering
\vspace{-2mm}
\caption{BConv Evaluation (w/ vs. w/o BAT, unit: $\mu s$)}\label{tab:bc_skew_shape}
\scriptsize
\vspace{-2mm}
\begin{tabular}{ccccccc}\hline
limb\_in $l$ &limb\_out $l'$ & Degree $N$ &Baseline &BAT &Speedup \\\hline
12 &28 &65536 &815.28 &135.91 & \textbf{6.00$\times$} \\
12 &36 &65536 &1054.89 &147.28 & \textbf{7.16$\times$} \\
16 &40 &65536 &165.18 &65.77 & \textbf{2.51$\times$} \\
24 &56 &65536 &318.92 &94.67 & \textbf{3.37$\times$} \\
\hline
\end{tabular}
\vspace{-2mm}
\end{table}


\subsubsection{MAT + BAT Evaluation -- NTT}  
\begin{figure}[h]
    \centering
    \vspace{-5mm}
    \subfloat[NTT/s vs TensorFHE+(A100) \label{fig:performance_compared_tensorFHE}]{{\includegraphics[width=0.485\columnwidth]{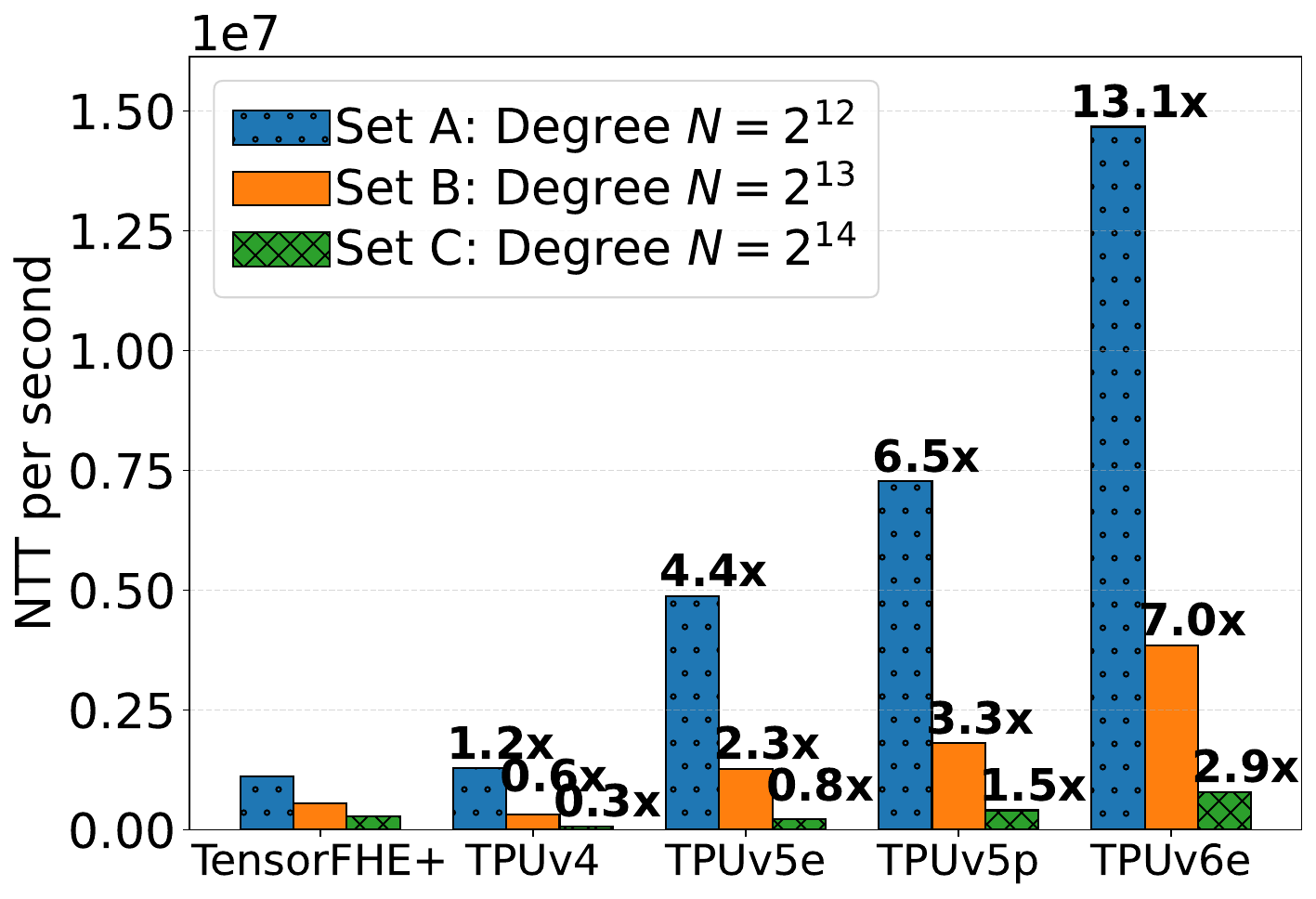}}}
    \subfloat[Impact of Batch Size.\label{fig:batch_effect}]{{\includegraphics[width=0.46\columnwidth]{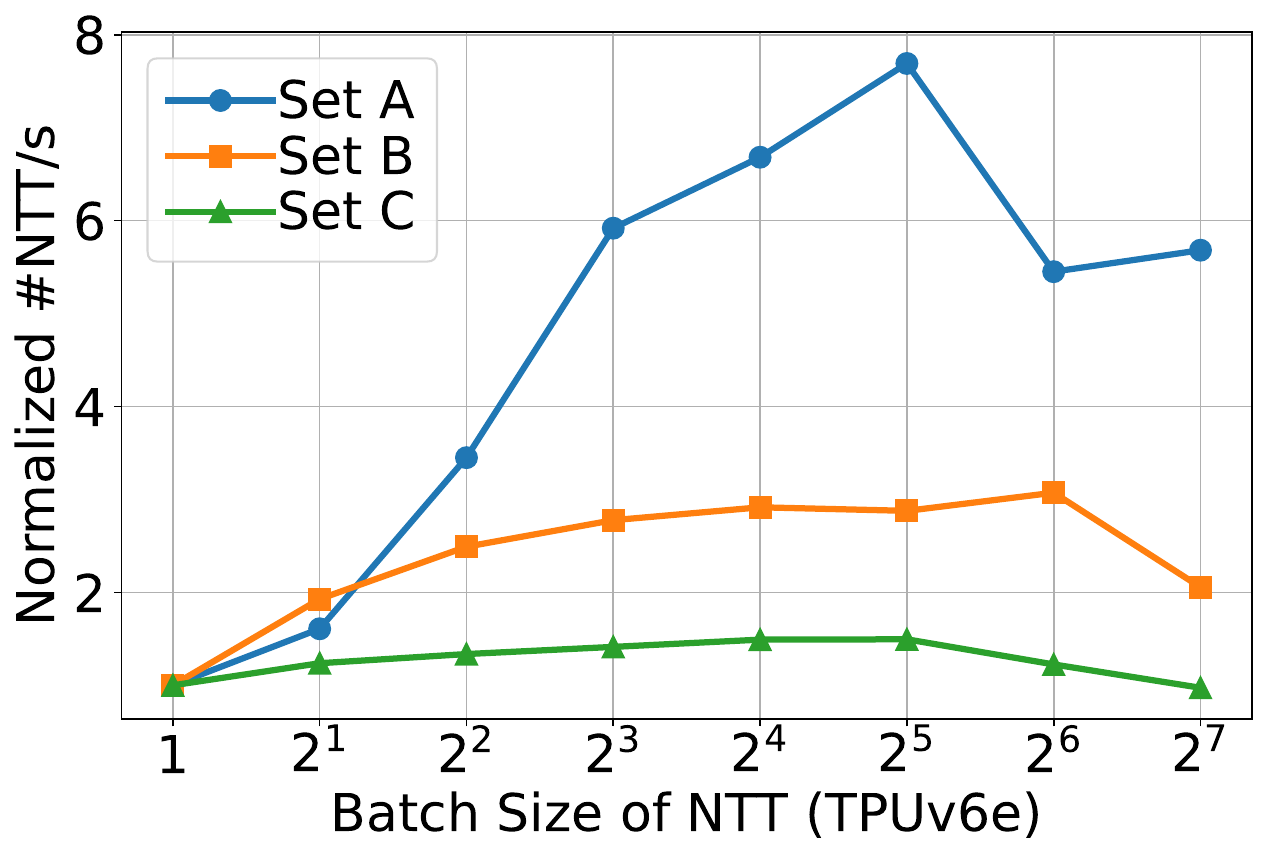}}}
    \vspace{-5mm}
    \caption{Ablation study: Impact of different hardware and batch size on NTT throughput. Parameters defined in \tabref{tab:ntt_throughput_setup}. }
    \label{fig:on_chip_mem}
    \vspace{-4mm}
\end{figure}

NTT is the dominant latency kernel in HE serving, so accelerating NTT directly improves overall performance. Under degree $N=(2^{12},2^{13},2^{14})$ with $\log_2Q=(109, 218, 438)$ (\tabref{tab:ntt_throughput_setup}), CROSS enables TPU v6e-8 to achieve up-to 99$\times$, 4$\times$, 2$\times$, 13.1$\times$, 1.2$\times$ higher throughput over HEAX,  FAB (U280 FPGA), HEAP (U280),  TensorFHE+ (A100), WarpDrive (A100) in \figref{fig:performance_compared_tensorFHE}. The less throughput improvement at higher degree is caused by a higher growth in overall computation from layout invariant NTT at $O(N^{3/2})$ compared with $O(N\log_2N)$ for others. \RevD{\textit{\textbf{Takeaway: }}This performance gain is attributed to two key optimizations in CROSS: (1) \textit{BAT} reduces redundant compute over zeros and its memory consumption, one type conversion, and halves temporal reduction length. (2) \textit{MAT} eliminates runtime transpose and shuffling by embedding them into computation.}

TPUs favor large batch as it reuses common parameters to save off-chip memory access before it overflows on-chip memory, as shown in \figref{fig:batch_effect}. NTT with higher degree benefits less from batching and achieves peak throughput at smaller batch size. This is because it might introduce back-and-forth data access to off-chip HBM, when multi-batch data of high-degree ciphertexts exceed on-chip memory. The optimal batch size for one tensor core of TPUv6e under Set A/B/C/D is 32/16/16/8, giving 7.7${\times}$/2.9${\times}$/1.5${\times}$/1.4${\times}$ throughput improvement.

\begin{table}[!h]\centering
\vspace{-2mm}
\caption{NTT Throughput (\#KNTT/s) Evaluation}\label{tab:ntt_throughput}
\vspace{-2mm}
\resizebox{0.48\textwidth}{!}{ 
\begin{tabular}{c|cc|cccc}\hline
Degree & TensorFHE+ & WarpDrive & v4-4 & v5e-4 &  v5p-4 & v6e-8   \\\hline
$N=2^{12}$ &1116 &12181 &1284 &4878 &7274 &14668 \\
$N=2^{13}$ &546 &4675 &323 &1276 &1812 &3850 \\
$N=2^{14}$ &276 &2088 &75 &223 &407 &793 \\
\hline
\end{tabular}}
\vspace{-2mm}
\end{table}

\begin{table}[!htp]
\centering
\vspace{-1mm}
\caption{Latency and energy efficiency of HE kernels (default $N=2^{16}$). Latency: $\mu s$. 
Values of baseline (in gray) come from their original paper.
The TPU configuration consuming similar power as each baseline is highlighted in green.
Speedup is ratio of gray to green using the configuration setup of the baseline.
Takeaway: CROSS enables TPUv6e to obtain lower amortized single-batch latency and SoTA energy efficiency.}
\label{tab:perf_cmp}
\footnotesize
\vspace{-2.4mm}
\resizebox{0.48\textwidth}{!}{
\begin{tabular}{c|ccccc}\hline
\textbf{Library} & \textbf{$L, \log_2q, Dnum$} &\textbf{ HE-Add} & \textbf{HE-Mult} & \textbf{Rescale} & \textbf{Rotate} \\ \hline
\rowcolor[gray]{0.9}\textbf{FIDESlib (RTX4090)}\cite{agullódomingo2025fideslibfullyfledgedopensourcefhe} 
& $30,59,3$ & \textbf{51}  & \textbf{1084} & \textbf{156} & \textbf{1107} \\
\textbf{CROSS (v4-8) }     &  \multirow{5}{*}{$60,28,3$}  & \textbf{13.1}  & \textbf{7336} & \textbf{1195} & \textbf{4491} \\ 
\textbf{CROSS (v5e-4) }     &   & \textbf{17.6}  & \textbf{2201} & \textbf{309} & \textbf{1418} \\ 
\textbf{CROSS (v5p-8) }     &   & \textbf{6.5}    & \textbf{1101} & \textbf{158} & \textbf{993} \\ 
\textbf{CROSS (v6e-4) }     &   & \textbf{8.2}    & \textbf{1364} & \textbf{187} & \textbf{941} \\
\rowcolor[HTML]{E6EFDB}\textbf{CROSS (v6e-8) }     &   & \textbf{4.0}  & \textbf{697} & \textbf{95} & \textbf{496} \\ \hdashline
\rowcolor[gray]{0.9}\textbf{Cheddar (RTX4090)}\cite{kim2024cheddar} 
& $48,\leq31,12$ & \textbf{48} & \textbf{533} & \textbf{68} & \textbf{476} \\ 
\textbf{CROSS (v4-8) }     &  \multirow{5}{*}{$48,28,3$} & \textbf{15.5}  & \textbf{5652} & \textbf{969} & \textbf{3799} \\ 
\textbf{CROSS (v5e-4) }     &   & \textbf{13.6}    & \textbf{1480} & \textbf{236} & \textbf{1161} \\ 
\textbf{CROSS (v5p-8) }     &   & \textbf{5.6}    & \textbf{851} & \textbf{143} & \textbf{769} \\ 
\textbf{CROSS (v6e-4) }     &   & \textbf{6.5}    & \textbf{951} & \textbf{145} & \textbf{787} \\
\rowcolor[HTML]{E6EFDB}\textbf{CROSS (v6e-8) }     &   & \textbf{3.5}    & \textbf{487} & \textbf{74} & \textbf{393} \\ \hdashline
\rowcolor[gray]{0.9}\textbf{FAB (U280)}\cite{agrawal2022fab} 
& $32,52,4$  & \textbf{40} & \textbf{1710} & \textbf{190} & \textbf{1570} \\ 
\textbf{CROSS (v4-8) }     & \multirow{5}{*}{$64,28,4$}  & \textbf{18.6}    & \textbf{7992} & \textbf{1260} & \textbf{5253} \\ 
\textbf{CROSS (v5e-4) }     &   & \textbf{19.0}  & \textbf{2238} & \textbf{344} & \textbf{1723} \\ 
\textbf{CROSS (v5p-8) }     &   & \textbf{5.9}    & \textbf{1190} & \textbf{173} & \textbf{1098} \\ 
\rowcolor[HTML]{E6EFDB} \textbf{CROSS (v6e-4) }     &   & \textbf{8.8}  & \textbf{1414} & \textbf{194} & \textbf{1080} \\
\textbf{CROSS (v6e-8) }     &   & \textbf{6.3}    & \textbf{709} & \textbf{97} & \textbf{547} \\ \hdashline
\rowcolor[gray]{0.9}\textbf{HEAP (8${\times}$U280)}\cite{heap} 
&  $N=2^{13},log_2Q=216$  & \textbf{1} & \textbf{28} & \textbf{10} & \textbf{25} \\ 
\textbf{CROSS (v4-8) }     & \multirow{5}{*}{\makecell{Set B \\ $8,28,3$}}  & \textbf{18.2}    & \textbf{197.7} & \textbf{84.6} & \textbf{241.7} \\ 
\textbf{CROSS (v5e-4) }     &   & \textbf{2.5}  & \textbf{34.0} & \textbf{5.1} & \textbf{40.5} \\ 
\textbf{CROSS (v5p-8) }     &   & \textbf{7.8}    & \textbf{48.0} & \textbf{17.3} & \textbf{77.4} \\ 
\textbf{CROSS (v6e-4)}    &   & \textbf{3.2}  & \textbf{20.7} & \textbf{17.3} & \textbf{31.0} \\
\rowcolor[HTML]{E6EFDB} \textbf{CROSS (v6e-8) }     &   & \textbf{6.5}    & \textbf{12.7} & \textbf{11.2} & \textbf{15.9} \\ \hdashline
\rowcolor[gray]{0.9}\textbf{BASALISC (ASIC)}\cite{basalisc_chip} 
& $32,40,3$ & \textbf{8} & \textbf{312} & \textbf{N/A} & \textbf{313} \\ 
\textbf{CROSS (v4-8) }     & \multirow{5}{*}{$47,28,3$} & \textbf{14.9}  & \textbf{5825} & \textbf{919} & \textbf{3659} \\ 
\textbf{CROSS (v5e-4) }     &   & \textbf{14.5}  & \textbf{1559} & \textbf{228} & \textbf{1091} \\ 
\textbf{CROSS (v5p-8) }     &   & \textbf{5.7}    & \textbf{1072} & \textbf{137} & \textbf{818} \\ 
\rowcolor[HTML]{E6EFDB} \textbf{CROSS (v6e-4) }     &   & \textbf{6.6}  & \textbf{955} & \textbf{135} & \textbf{754} \\
\textbf{CROSS (v6e-8) }     &   & \textbf{3.6}  & \textbf{488} & \textbf{67} & \textbf{328} \\ \hdashline
\rowcolor[gray]{0.9}\textbf{WarpDrive (A100)}\cite{warpdrive} 
& $34,28,$ Not Known & \textbf{61} & \textbf{4284} & \textbf{241} & \textbf{5659} \\ 
\textbf{CROSS (v4-8) }     & \multirow{5}{*}{$36,28,3$}  & \textbf{42.8}  & \textbf{4373} & \textbf{724} & \textbf{2875} \\ 
\textbf{CROSS (v5e-4) }     &   & \textbf{14.0}  & \textbf{1114} & \textbf{175} & \textbf{908} \\ 
\textbf{CROSS (v5p-8) }     &   & \textbf{11.0}   & \textbf{656} & \textbf{117} & \textbf{687} \\ 
\rowcolor[HTML]{E6EFDB} \textbf{CROSS (v6e-4) }     &   & \textbf{10.9}  & \textbf{714} & \textbf{106} & \textbf{593} \\
\textbf{CROSS (v6e-8) }     &   & \textbf{5.0}  & \textbf{358} & \textbf{61}& \textbf{307} \\ \hdashline
\rowcolor[gray]{0.9}\textbf{CraterLake (ASIC)}\cite{CraterLake} &  $51,28,3$ & \textbf{9} & \textbf{35} & \textbf{9} & \textbf{27} \\
\rowcolor[gray]{0.9} \textbf{OpenFHE (AMD 9950X3D)} & $51,28,3$ & \textbf{15390} & \textbf{417651} & \textbf{22670} & \textbf{397798} \\ 
CROSS (AMD 9950X3D)  &  \multirow{6}{*}{\makecell{ \\ \textbf{Default} \\ Set D \\ \textbf{51,28,3}}} & 79  & {132472} & {27281} & {72741} \\
\textbf{CROSS (v4-8) }     &   & \textbf{15.8}  & \textbf{5962} & \textbf{988} & \textbf{3975} \\ 
\textbf{CROSS (v5e-4) }     &   & \textbf{15.0}  & \textbf{1574} & \textbf{242} & \textbf{1149} \\ 
\textbf{CROSS (v5p-8) }     &   & \textbf{11.0}   & \textbf{782} & \textbf{131} & \textbf{853} \\ 
\rowcolor[HTML]{E6EFDB} \textbf{CROSS (v6e-4) }     & $51,28,3$  & \textbf{6.8}  & \textbf{1007} & \textbf{149} & \textbf{798} \\
\textbf{CROSS (v6e-8) }     &   & \textbf{3.5}  & \textbf{509} & \textbf{77} & \textbf{414} \\ \hline \hline
\textbf{Energy Efficiency}  & vs OpenFHE & $\mathbf{2253\times\uparrow}$ & $\mathbf{415\times\uparrow}$ & $\mathbf{152\times\uparrow}$ & $\mathbf{498\times\uparrow}$ \\ 
\textbf{Improvement} & vs FIDESlib & $\mathbf{12.8\times\uparrow}$ & $\mathbf{1.55\times\uparrow}$ & $\mathbf{1.64\times\uparrow}$ & $\mathbf{2.23\times\uparrow}$ \\ 
Over & vs WarpDrive & $\mathbf{5.61\times\uparrow}$ & $\mathbf{6.00\times\uparrow}$ & $\mathbf{2.27\times\uparrow}$ & $\mathbf{9.54\times\uparrow}$ \\ 
\textbf{Publicly} & vs Cheddar & $\mathbf{13.6\times\uparrow}$ & $\mathbf{1.10\times\uparrow}$ & $\mathbf{0.92\times}$ & $\mathbf{1.21\times\uparrow}$ \\ 
\textbf{Available }& vs FAB & $\mathbf{4.55\times\uparrow}$ & $\mathbf{1.21\times\uparrow}$ & $\mathbf{0.98\times}$ & $\mathbf{1.45\times\uparrow}$ \\
\textbf{ Devices}& vs HEAP & $\mathbf{0.15\times}$ & $\mathbf{2.20\times\uparrow}$ & $\mathbf{0.89\times}$ & $\mathbf{1.58\times\uparrow}$ \\ \hdashline
Over& vs BASALISC & $\mathbf{1.20\times\uparrow}$ & $\mathbf{0.33\times}$ & \textbf{N/A} & $\mathbf{0.42\times}$ \\ 
\textbf{Unavailable Devices} & vs CraterLake & $\mathbf{1.32\times\uparrow}$ & $\mathbf{0.03\times}$ & $\mathbf{0.06\times}$ & $\mathbf{0.03\times}$ \\ 
\hline
\end{tabular}}
\vspace{-3mm}
\end{table}

\insertFigure{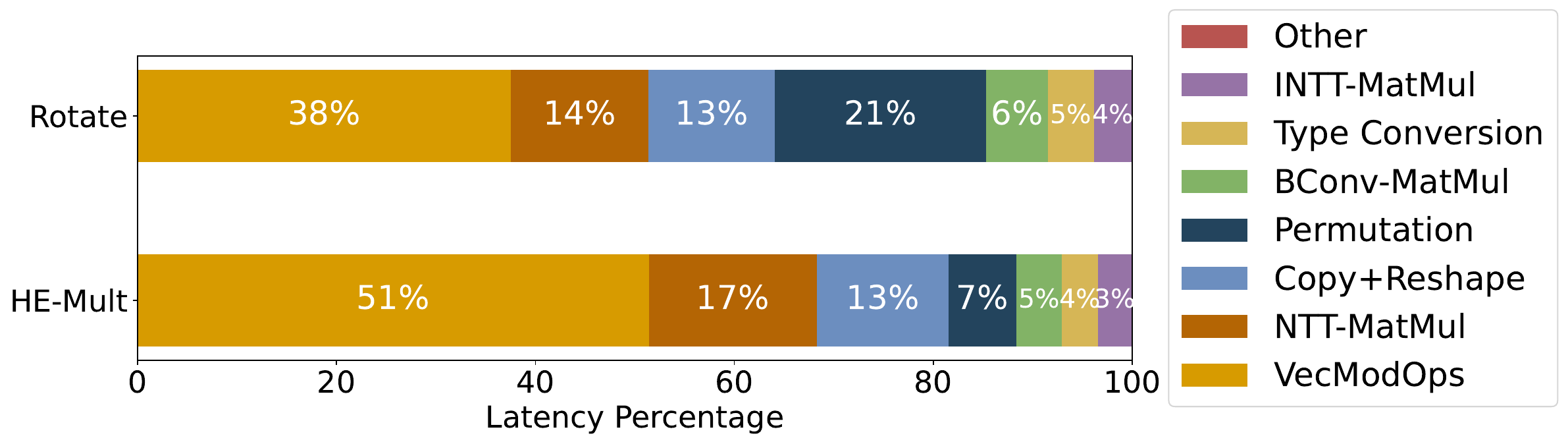}{Latency breakdown of HE multiplication and rotation on TPUv6e, under Set D from \tabref{tab:perf_cmp}.}

\subsection{Evaluating Performance of HE Kernels}
We scale TPUs to achieve roughly the same power as other devices in prior works and compare amortized single-batch latency of HE kernels in \tabref{tab:perf_cmp}, adopting the security parameters that yielded the peak performance reported in original publications of prior works or the corresponding double rescaling version with twice number of 32-bit moduli.

\paragraph{CROSS vs SoTAs} After scaling Tensor Cores to match the power envelope of prior systems, CROSS on a single TPUv6e VM achieves significant speedups for HE-Mult/Rotate: $415\times$/$498\times$ over OpenFHE, $1.55\times$/$2.23\times$ over FIDESlib, $1.21\times$/$1.45\times$ over FAB, $2.2\times$/$1.58\times$ over HEAP, and $6\times$/$9.54\times$ over WarpDrive. Compared to Cheddar \cite{kim2024cheddar}, the SoTA GPU library, we utilize a configuration ($dnum=3$, 65 total moduli) that incurs $1.22\times$ higher memory consumption than the baseline ($dnum=12$, 53 total moduli). Despite this overhead, CROSS achieves speedup performance in HE-Mult ($1.1\times$) and Rotate ($1.21\times$). On average, CROSS/TPUv6e delivers $451\times$, $7.81\times$, $1.83\times$, $1.31\times$, $1.86\times$, and $1.15\times$ higher energy efficiency than OpenFHE, WarpDrive, FIDESlib, FAB, HEAP, and Cheddar, respectively. Even against non-public specialized ASICs, CROSS outperforms BASALISC ($1.20\times$) and CraterLake ($1.32\times$) in HE-Add efficiency, while narrowing the HE-Mult performance gap to $3\times$, $33\times$ and Rotate gap to 2.4${\times}$, 33$\times$. These establish CROSS on TPUv6e as SoTA energy-efficiency frontier for HE operators in commodity devices.


\paragraph{CROSS for CPU} On the AMD 9950X3D, CROSS delivers up to \textbf{5.46{$\times$}/3.15{$\times$} speedup} in HE-Mult/Rotate over OpenFHE. The improvement stems from offloading partial computation of redundant zeros and runtime data reordering to offline (compile time). Note that CROSS enables CPU to run faster by switching from $O(N\log_2N)$ complexity radix-2 cooley-tukey NTT algorithm into $O(N\sqrt{N})$ layout invariant 3-step NTT algorithm, highlighting the need to unleash modern hardware's efficiency by performing workload at coarse granularity of vector or matrix instead of individual elements.

\paragraph{Effects of Security Parameters} Increasing (1) the total number of limbs or (2) the digit number (\textit{dnum}), which specifies the number of partitions used for digit decomposition in hybrid key switching~\cite{hybrid_keyswitch}, could both increase the required computation, leading to longer latency on TPU (\tabref{tab:perf_cmp}).

\paragraph{Latency Breakdown} \figref{fig:TPUv6e_Profiled_Latency} provides a runtime breakdown for HE-Mult and Rotate on a single TPUv6e Tensor Core (Set D), identifying key execution bottlenecks.
 
\noindent $\bullet$ \textbf{HE-Mult}: Matrix multiplications in NTT/INTT/BConv dominate arithmetic complexity but only contribute 25\% of latency. VecModMul instead takes 51\%, making HE-Mult VPU-bound. The remaining 13\% + 7\% is XLA-induced memory re-layout to (8,128) tiles to better utilize VReg, not algorithmic cost. BAT adds 4\% extra type conversion to change 32-bit data into bytes, as it might trigger layout conversion.

\noindent $\bullet$ \textbf{Rotate}: The same VPU-bounded bottleneck shows again, dominating 38\% latency while MatMuls from NTT, INTT and BConv take only 24\% latency, highlighting the efficiency and speed of MXU. The 21\% permutation represents worst-case cost, which introduces random gather/scatter when MAT cannot embed a given permutation pattern into computation.


\textbf{Takeaway}: TPUv6e achieves SoTA energy efficiency among commodity devices. Both HE-Mult and Rotate are bounded by vectorized operation with Rotate further bounded by permutation, motivating future work to optimize algorithms to further embed these algorithmic permutation into computation and to convert vectorized operations into low-precision MatMul to be accelerated by high-throughput MXU. Moreover, CROSS's optimizations generalize to any architecture with dedicated matrix or vectorized engine.

\subsection{Evaluating Performance of HE ML Workload}
\label{q:re_q1}
\paragraph{MNIST} We evaluate CROSS on a convolution network ($2\times\{\text{Conv-ReLU-AvgPool}\} \to \text{FC} \to \text{ReLU} \to \text{FC}$)~\cite{wise_HE_winograd} using a batch size of 64 with $3\times32\times32$ MNIST images~\cite{brutzkus2019lowlatencyprivacypreserving}. The HE parameters are set to $N=2^{13}$, $\text{dnum}=3$, $L=18$, and $\log_2 q=28$ without bootstrapping. Weights are encoded as plaintexts while inputs are encrypted as ciphertexts. On TPUv6e-8, CROSS achieves an amortized inference latency of 270~ms per image, a 10$\times$ speedup over Orion~\cite{orion_nyu} with the same 98\% accuracy. It's because BAT and MAT optimizations are mathematically lossless transformations that preserve the exact computational results of the modular arithmetic.  


\paragraph{Logistic Regression} CROSS uses one TPUv6e tensor core to achieve 84 ms for a single iteration of LR, achieving 1.06$\times$ throughput per watt than Cheddar (RTX 4090).

\textit{\textbf{Takeaway:}} We show that TPUv6e as an AI ASIC can be a promising platform for accelerating HE ML workloads. It favors lower precision moduli, giving better energy efficiency for HE ML workloads without bootstrapping.

\subsection{Evaluating Packed Bootstrapping}
CROSS adopts pack bootstrapping~\cite{agrawal2023mad} under default Set D from \tabref{tab:ntt_throughput_setup}. CROSS on TPUv6e-8 achieves $1.5\times$ throughput (measured by number of bootstraps-per-second) over the SoTA GPU library Cheddar~\cite{fan2022tensorfhe,warpdrive, tensorfhe_plus} on an NVIDIA RTX 4090 GPU, validating the efficacy of redundancy reduction in BAT and layout reordering elimination in MAT. CROSS enables high compute utilization for HE kernels of the bootstrapping, making TPUv6e the SotA throughput machine for workloads built on HE operators with bootstrapping too.

CROSS on TPUv6e shows 5${\times}$ bootstrapping throughput gap to  HE ASIC, CraterLake\cite{CraterLake}, when scaling tensor cores to consume roughly same power. Software reasons are:

$\bullet$ \textbf{Limited Inter-Kernel Optimization}: CROSS does not explore pipelining and fusion between sequential vectorized operators and matrix multiplication of HE kernels of bootstrapping. Consequently, intermediate results are written back to HBM, incurring back-and-forth memory access. 

$\bullet$ \textbf{Inefficient Permutation in Automorphism}: CROSS's MAT could embed all reordering of NTT but not all permutations in Automorphism into computation. Therefore, automorphism is mapped as random scatter and gather of many degree-length vectors, causing layout transformations. How to embed arbitrary permutation into computation remains an open-question.

\begin{table}[!t]\centering
\vspace{2mm}
\caption{Packed Bootstrapping (v6e-8's Speedup/breakdown)}\label{tab:bootstraping}
\vspace{-2mm}
\scriptsize
\resizebox{0.47\textwidth}{!}{
\begin{tabular}{cccccccc}\hline
Work  &FIDESlib &Cheddar &CL & v4-8  & v5e-4 & v5p-8 & v6e-8  \\\hline
Latency (ms)  & 169 & 31.6 &3.91 & 129.8 & 59.2 & 68.3 & 21.5\\
v6e's Speedup & \textbf{7.9$\times$} & \textbf{1.5$\times$} & \textbf{0.2$\times$} & \textbf{6$\times$} & \textbf{2.8$\times$} & \textbf{3.2$\times$} & 1\\
\end{tabular}}
\resizebox{0.47\textwidth}{!}{
\begin{tabular}{cccccc}\hline
v6e-8 Breakdown  &Automorphism  &VecModMul &(I)NTT &VecModAdd  &BConv  \\\hline
Lat. Ratio  &35.64\% &25.55\% &16.87\% &15.29\% &6.65\%  \\
\hline
\end{tabular}}
\vspace{-6mm}
\end{table}

\subsection{Ablation Study}

\subsubsection{Impact of Batch Size}
As shown in \figref{fig:batch_effect}, increasing batch size improves NTT throughput on TPUs by enabling reuse of shared parameters across ciphertexts. Such parameters include twiddle factors, evaluation keys, and CRT primes for basis conversion. This reuse reduces redundant off-chip memory traffic and increases operational intensity, shifting execution from a memory-bound towards a compute-bound regime. However, excessively large batches exceed on-chip memory capacity, causing contention between input coefficients and parameters and triggering repeated off-chip accesses, which degrades performance. Increasing on-chip memory capacity would give higher benefits at larger batch sizes. Under Set D, we observe that e-class TPUs (e.g., TPUv5e and TPUv6e) achieve peak performance with single-batch HE operators, whereas p-class TPUs (e.g., TPUv4 and TPUv5p) continue to benefit from amortized latency reductions at batch sizes of 2 or 4 because of large off-chip and on-chip memory.

\begin{figure}[ht]
    \centering
    \vspace{-4mm}
    \setlength{\fboxrule}{2pt}  
    \subfloat[VecModMul w/ diff. mod red.\label{fig:vecmodmul_diff_modred}]{{\includegraphics[width=0.48\columnwidth]{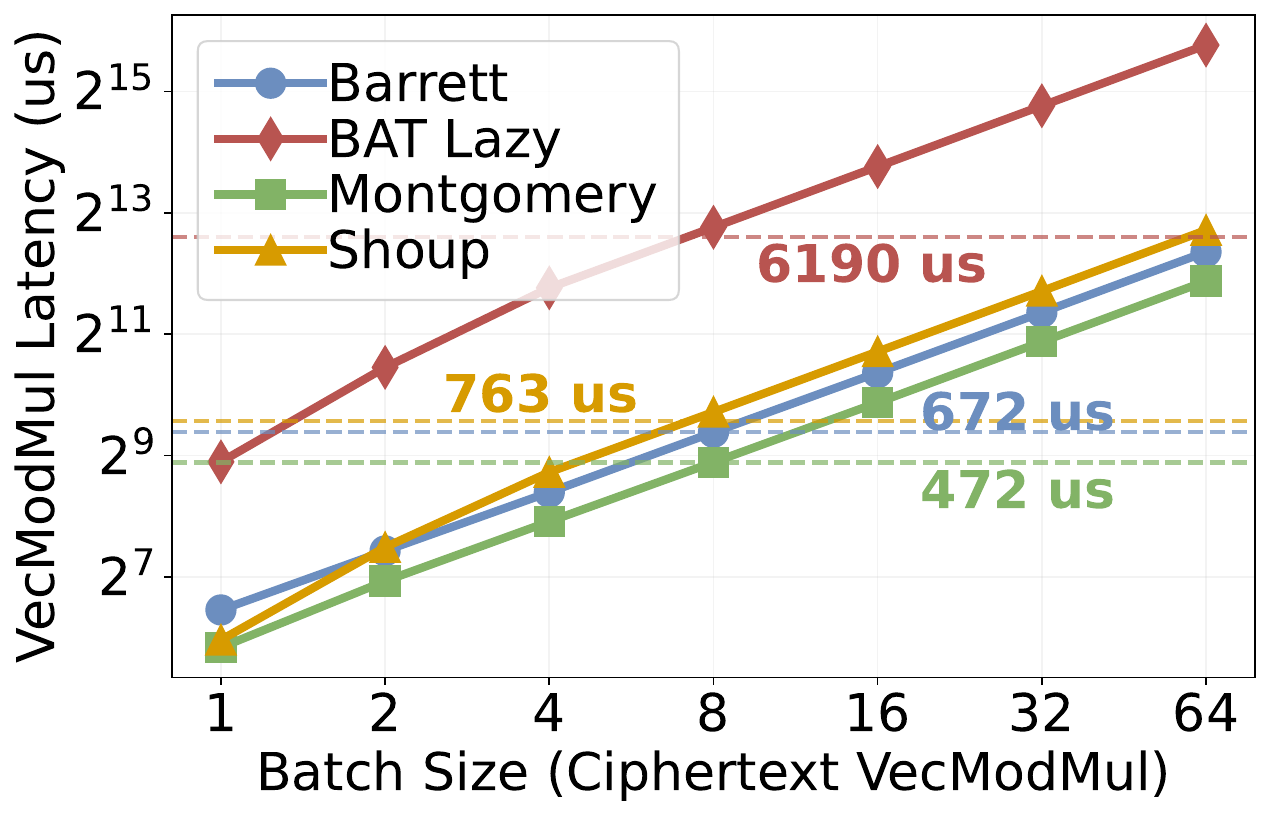}}}
    \subfloat[NTT w/ diff. mod red. \label{fig:ntt_diff_modred}]{{\includegraphics[width=0.48\columnwidth]{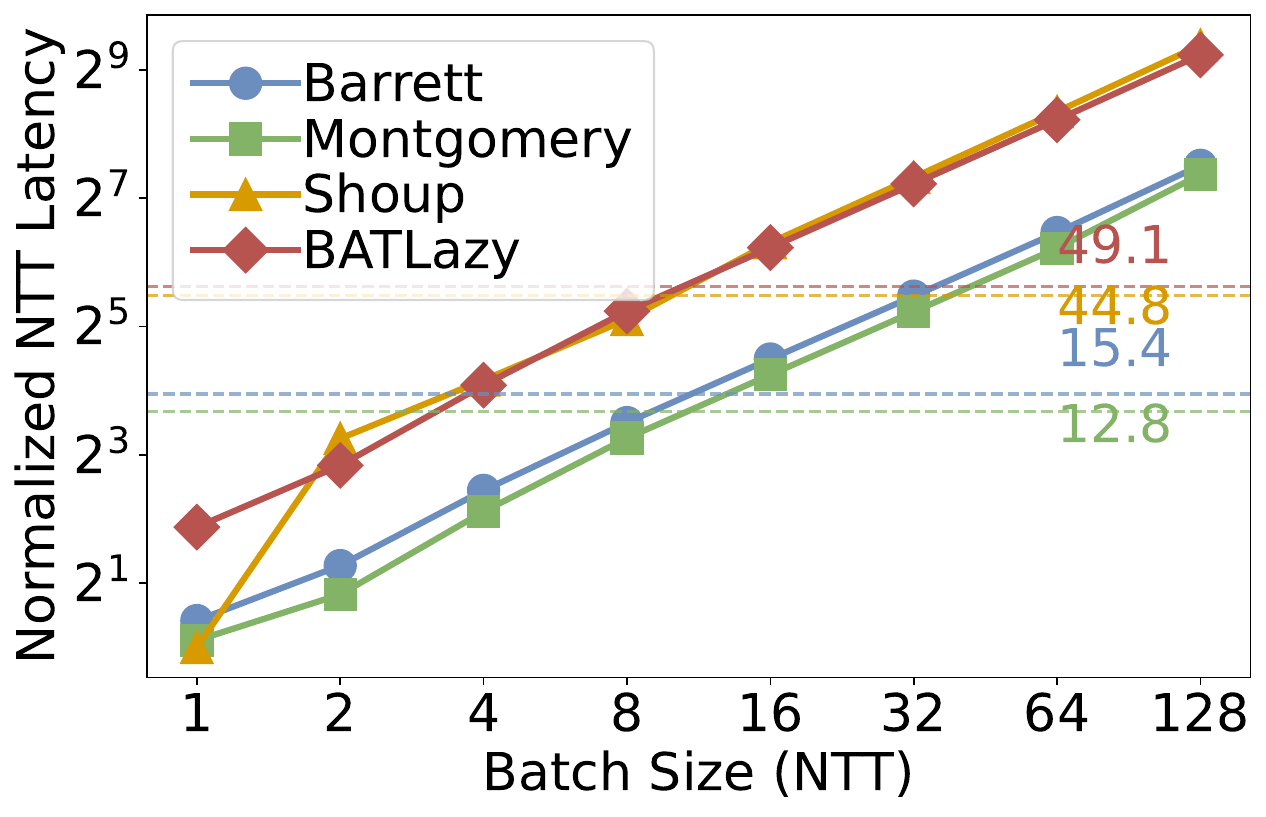}}}
    \vspace{-2mm}
    \caption{Ablation study: Impact of modular reduction (modred) on latency of NTT and VecModMul (Set D).}
    \label{fig:diff_modred}
    \vspace{-6mm}
\end{figure}

\subsubsection{Impact of Modular Reduction Algorithm}
\label{q:rf_q3}

Since VecModMul and NTT dominate HE compute latency, we evaluate three modular reduction algorithms, including \textit{Barrett}~\cite{Barrett}, \textit{Montgomery}~\cite{montgomery1985modular}, and \textit{Shoup}~\cite{shoup2001ntl}, across varying batch sizes (\figref{fig:diff_modred}). Our results indicate that Montgomery reduction is optimal for TPUv6e for both VecModMul and ModMatMul.

\noindent$\bullet$ \textbf{Setup:} We apply the BAT to Barrett and Montgomery reductions to optimize ModMatMul while apply reduction algorithms directly for VecModMul. As Shoup’s reduction relies on precompiled parameters that are incompatible with BAT, we implement it using the SoTA GPU high-precision scalar multiplication flow shown in \figref{fig:CROSS_Scalar_Flow}. We also evaluate ``BAT lazy reduction'', for which we apply BAT to reformulate modular reduction as low-precision matrix multiplication (details in \secref{sec:bat_lazy}). All experiments use a single TPUv6e Tensor Core under security parameter Set D.

\noindent$\bullet$ \textbf{VecModMul:}
Montgomery reduction achieves a 1.42$\times$ geomean speedup over Barrett across all batch sizes by reducing runtime computation. While Shoup's algorithm has the lowest theoretical arithmetic complexity, it needs 64-bit multiplication, hence being slower than Montgomery. BAT lazy reduction performs poorly as its reduction dimension is limited to the number of bytes in the data, i.e. $K=4$ for 32-bit data, resulting in inefficient MXU utilization.

\noindent$\bullet$ \textbf{NTT:}
NTT composes \textit{ModMatMul} and \textit{VecModMul} (\figref{fig:NTT_overall_flow}). \RevF{Crucially, the throughput gains from BAT-optimized MatMul further magnify gap between Montgomery and Shoup.} The only exception is the single-batch configuration, where the NTT kernel becomes memory-bound on TPUv6e, masking the computational advantages of Montgomery and Barrett.

\subsection{Unveiling Performance Gap to Dedicated FHE ASICs}
\label{sec:gap_analysis}

\textbf{HE ASIC SoTA:} Under HE-Mult / Rotate / bootstrapping as the workload, compared against HE ASIC accelerators CraterLake~\cite{CraterLake}, TPUv6e with \compiler is $33.2\times$/$33.2\times$/$5\times$ slower. The performance gap between CROSS on TPU and FHE ASICs is driven by three factors from hardware side:

\squishlist
 \item \textbf{Hardware-Friendly Moduli:}
Dedicated FHE ASICs~\cite{CraterLake} use fixed moduli (e.g., $2^{32}-v$, with $v$ as a 16-bit value) optimized for hardware. In contrast, CROSS supports arbitrary moduli, which can incur a 2$\sim$3$\times$ performance penalty depending on the implementation. 
\item \textbf{Low-Cost All-to-All Shuffling:}
Dedicated accelerators feature efficient shuffling mechanisms, such as the layout transpose unit in CraterLake~\cite{CraterLake} and the all-to-all connected NoC in FAB~\cite{agrawal2022fab}, which enable a butterfly recursive NTT with $O(N\log_2 N)$ complexity. This design choice yields up to a 16$\times$ performance advantage over layout invariant 3-step NTT with $O(N\sqrt{N})$ computational complexity when picking polynomial degree $N=2^{16}$.
\item \textbf{Extensive Compute and Large On-Chip Memory:}
Dedicated FHE ASICs allocate a significant portion of chip area to on-chip memory (e.g., 256 MB, double of TPUv4), which supports larger batch sizes and increased data reuse, thereby enhancing overall performance.
\squishend

All three factors explain $33\times$ latency gap in \tabref{tab:perf_cmp}.




\section{Related Work}
\label{sec:relatedWork}

Previous studies on hardware-accelerated HE follow two main tracks. The first track proposes custom ASIC designs with the sole purpose of accelerating HE workloads. Initial proposals were relatively small~\cite{f1} but quickly grew to chips requiring hundreds of MB in memory and hundreds of $\text{mm}^2$ in area~\cite{CraterLake, BTS, shivdikar2023gme,mz_micro_FHE}. While these designs achieve significant performance gains over CPU baselines, they would cost millions of dollars to fabricate and deploy. 

This motivates the second track of prior works, which use existing hardware to accelerate HE workloads. Following the path of AI acceleration, these works focus on GPU acceleration~\cite{fast_HE_GPU, fan2022tensorfhe,jung2021over, zhai2021acceleratingencryptedcomputingintel,shivdikar2022acceleratingpolynomialmultiplicationhomomorphic,shivdikar2023gme,GPU_HE_Accel,Kim_2020,kim2024cheddar, dathathri2018chet, GPU_WhitePaper_mention_21, FxHENN_HPCA23,IntelHEXL,li2025catgpuacceleratedfheframework, phantom_fhe, agullódomingo2025fideslibfullyfledgedopensourcefhe, wang2024chameleonefficientfhescheme,HE_engine,HE_Booster} as well as employing FPGAs~\cite{Sinha_FPGA_HE_HPCA,agrawal2022fab,HEAX20, CHAM_FPGA} as more configurable commodity hardware. Our work falls into this second category, proposing a novel compilation techniques to better utilize \emph{existing} hardware to accelerate HE workloads at no additional hardware cost. This work establishs the path of adapting AI accelerators for HE workloads to achieve better energy efficiency than SoTA GPUs and FPGAs solutions, enabling the same chip to support both privacy-preserving AI and AI.
\section{Conclusion}
\label{sec:conclusion}
This work introduces \compiler, the first compilation framework that enables AI accelerators, such as Google TPUs, to efficiently execute Homomorphic Encryption (HE) operators, achieving superior performance per watt compared to SoTA FPGA and GPU implementations. \compiler establishes new NTT throughput records previously held by GPUs at low polynomial degrees ($N \le 2^{12}$). More broadly, CROSS proposes compilation techniques to transform applications with statically scheduled modular arithmetic and deterministic permutations into TPU-friendly computational kernels, and hence allows HE workloads to directly inherit the energy efficiency and throughput of modern AI ASICs without any hardware modification. This result positions AI accelerators (e.g., TPUv6e) as a viable and energy-efficient throughput machine for privacy-preserving computation. We leave application-level optimizations and end-to-end benchmarks to future work.
\section{Acknowledge}
This work was supported in part by ACE, one of the seven centers in JUMP 2.0, a Semiconductor Research Corporation (SRC) program sponsored by DARPA. The TPU access of this work is supported by Google TPU Research Cloud. We thank Todd Austin, Mohit Tiwari, Manoj Kumar, Lok Yan, Baiyu Li, Shruthi Gorantala, Bryant Gipson, Ingrid Verbauwhede, Brandon Reagen, Moin Qureshi, Srini Devadas, Mattan Erez, Karthik Garimella, Suvinay Subramanian, Rashimi Agrawal, Scott Hrastar, Simon Langowski, Shannon Egan, Brock Dorion, and anonymous reviewers for insightful feedbacks.

\bibliographystyle{IEEEtranS}
\bibliography{refs}

\appendix 
\section{Artifact Appendix}

\subsection{Abstract}

{ \em We provide scripts to reproduce latency of BAT (\tabref{tab:bat_perform}) and BConv (\tabref{tab:bc_skew_shape}), throughput of NTT (\tabref{tab:ntt_throughput}, \figref{fig:performance_compared_tensorFHE} and \figref{fig:batch_effect}), latency of HE operators (\tabref{tab:perf_cmp}), performance of different modular reduction algorithms (\figref{fig:vecmodmul_diff_modred} and \figref{fig:ntt_diff_modred}), latency profiling (\figref{fig:TPUv6e_Profiled_Latency}), and packed bootstrapping estimation (\tabref{tab:bootstraping}). We provide individual script to run each experiment and obtain the final results.} From \secref{sec:he_kernels}, we provide necessary background, experimental setup, and additional results for precise results reproduction and easily adopting CROSS's contributions.

\subsection{Artifact check-list (meta-information)}

{\small
\begin{itemize}
  \item {\bf Run-time environment: } Python 3.13, jax[tpu]
  \item {\bf Hardware: } TPUv4, TPUv5e, TPUv5p, TPUv6e
  \item {\bf Experiments: } Critical experiments (\tabref{tab:bat_perform}, \tabref{tab:bc_skew_shape}, \figref{fig:performance_compared_tensorFHE}, \tabref{tab:perf_cmp}), Optional Experiment (\tabref{tab:bootstraping}, \figref{fig:vecmodmul_diff_modred}, \figref{fig:ntt_diff_modred}, \tabref{tab:ntt_throughput}, \figref{fig:TPUv6e_Profiled_Latency}).
  \item {\bf Publicly available?: } \url{https://github.com/EfficientPPML/CROSS}
  \item {\bf Code licenses (if publicly available)?: } MIT
  \item {\bf Archived (provide DOI)?: } \url{https://doi.org/10.5281/zenodo.18077699}
\end{itemize}
}

\subsection{Description and Installation}

The CROSS framework runs on any device supporting JAX. For reproducing the performance on TPU, the access of TPUv4,v5e,v5p,v6e is required. And it runs on Python 3.13 with following packages. We note that if running on GPU, the data type for 8-bit convolution should be changed from uint8 into int16 as GPU does not support uint8 based convolution.

\begin{lstlisting}
pip install -U "jax[tpu]", xprof, absl-py, pandas, gmpy2
\end{lstlisting}

\subsection{Experiment workflow}
Contact authors to obtain the access of TPUs, and then run each provided script to get results for individual experiments. The entire repository is organized by absltest, and profiled by XProf with the compiled kernel latency being printed out in the terminal and written out as a csv.

\subsection{Evaluation and expected results}
Values in the tables and figures are expected results, and the difference of profiled results should have $\leq\pm 5\%$ difference compared to the provided results in the paper.

\subsection{Compute Pattern of Bottleneck Kernels in SoTA HE Library}
\label{sec:he_kernels}
HE workloads fundamentally boil down to the scheduled invocation of essential HE operators, including HE Multiplication (HE-Mult), Relinearization (Relin.), Rotate, and Rescale. In other words, the performance of these operators directly determines the overall serving latency of HE workloads. 
To identify performance bottlenecks, we profiled SoTA HE algorithms used by HE FPGA~\cite{agrawal2022fab} and ASIC~\cite{sharp23} on an AMD Ryzen 9 5950X CPU with AVX support using OpenFHE library. The latency breakdown, presented in \figref{fig:CPU_Profiled_Latency}, reveals NTT, and its inverse (INTT), BConv, and Vectorized Modular Multiplication (\texttt{VecModMul}) and Addition (\texttt{VecModAdd}) are the five most time-consuming HE kernels in both CKKS and BFV schemes. 

\subsubsection{Radix-2 Cooley-Tukey NTT algorithm (\textbf{Butterfly NTT})}
\label{sec:ntt}

The NTT converts polynomial representations from the coefficient domain to the evaluation domain, where polynomial multiplication simplifies to element-wise (vectorized) coefficient multiplication. The NTT and INTT are computationally intensive, accounting for approximately 45.1\% to 86.3\% of the overall latency in various HE operators. 

The detailed algorithm is provided in Algorithm~\ref{alg:NTT}. In general, an $N$-point NTT consists of \(\log_2(N)\) stages. Compute-wise, each stage comprises \(\frac{N}{2}\) vectorized modular multiplications, additions, and subtractions, denoted as \(\frac{N}{2}\)-\texttt{VecModMul}, \(\frac{N}{2}\)-\texttt{VecModAdd}, and \(\frac{N}{2}\)-\texttt{VecModSub}, respectively; Memory-wise, each stage also requires bit-complement shuffling~\cite{dally2004principles}.

\begin{figure}[t]
    \centering
    \vspace{-3mm}
    \subfloat[8-input butterfly NTT \label{fig:NTT_diagram}]{{\includegraphics[width=0.37\columnwidth]{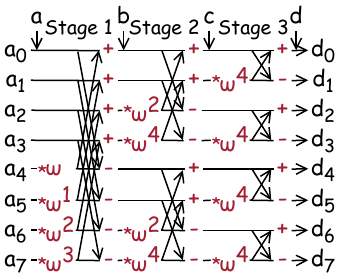}}}
     \subfloat[Basis Change (BConv) \label{fig:bc_two_steps}]{{\includegraphics[width=0.65\columnwidth]{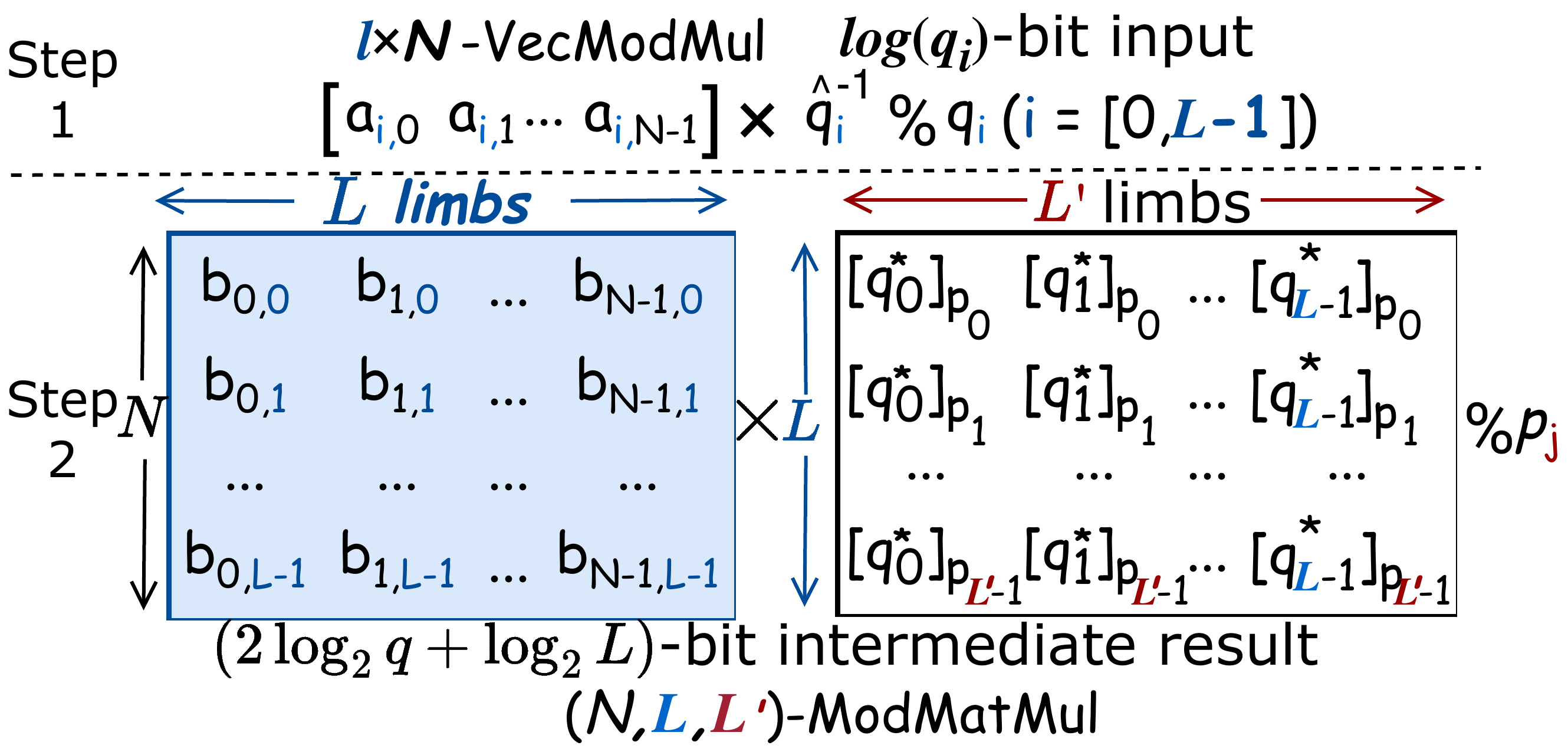}}}
    \vspace{-6mm}
    \caption{Illustration of computation patterns of butterfly NTT and basis change, superscript represents exponent.}
    \label{fig:cross_overall_perf}
    \vspace{-2mm}
\end{figure}

Considering 8-point NTT depicted in \figref{fig:NTT_diagram} as an example, in stage 1, we perform a 4-element vectorized modular multiplication between $[a_4, a_5, a_6, a_7]$ and $[\omega, \omega^2, \omega^3, \omega^4]$:
\begin{equation*}
[\tilde{a}_4, \tilde{a}_5, \tilde{a}_6, \tilde{a}_7] = [a_4 \cdot \omega, \ a_5 \cdot \omega^2, \ a_6 \cdot \omega^3, \ a_7 \cdot \omega^4] \mod q
\end{equation*}
Then $vec=[a_0, a_1, a_2, a_3, \tilde{a}_4, \tilde{a}_5, \tilde{a}_6, \tilde{a}_7]$ is being reduced with the bit complement shuffled result of itself $vec = vec \pm \text{bit\_complement\_shuffle}(vec)$ using vectorized modular addition and subtraction to produce:
\begin{equation*}
[b_0, b_1, b_2, b_3] {=} [a_0 {+} \tilde{a}_4, \ a_1 {+} \tilde{a}_5, \ a_2 {+} \tilde{a}_6, \ a_3 {+} \tilde{a}_7] \bmod q
\end{equation*}
\begin{equation*}
[b_4, b_5, b_6, b_7] {=} [a_0 {-} \tilde{a}_4, \ a_1 {-} \tilde{a}_5, \ a_2 {-} \tilde{a}_6, \ a_3 {-} \tilde{a}_7] \bmod q
\end{equation*}
The bit-complement shuffling in the first stage has a group size of 8 elements~\cite{dally2004principles}, as illustrated by the permutation arrows in \figref{fig:NTT_diagram}. Subsequent stages perform bit-complement permutation with the group sizes decreasing by 2 per stage.

\insertFigure{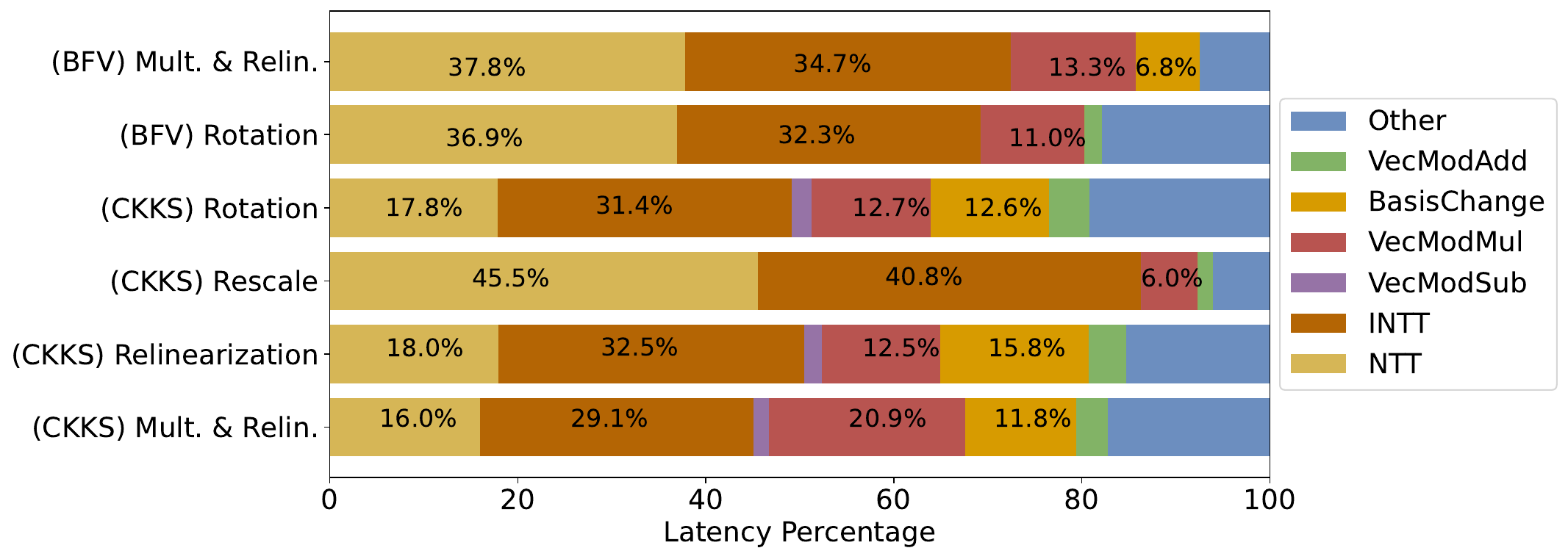}{Latency profiling of HE operators using OpenFHE, picking representative values for the parameters in \tabref{tab:Notation}. Vectorized Modular Multiplication (\texttt{VecModMul}), Addition (\texttt{VecModAdd}) and Subtraction (\texttt{VecModSub}) only count for latency not belonging to (I)NTT/BConv.} 
On TPUv4, radix-2 cooley-tukey NTT following its algorithmic processing order runs $\sim30\times$ slower than MAT-based NTT, even though radix-2 cooley-tukey NTT has lower computational complexity of $O(NlogN)$ than $O(N^{3/2})$ of MAT-based NTT, as shown in \tabref{tab:ntt_butterfly_vs_recursive}. This is because (1) heavy reordering and (2) zero MXU utilization.

\begin{algorithm}
\caption{Radix-2 Cooley-Tukey NTT~\cite{NTT}} \label{alg:NTT}
\begin{algorithmic}[1]
 \Require Coefficients of input polynomial $P_m=\left(a_0, a_1, \cdots, a_{N-1} \right)$, $N$-th root of unity $\omega$, total degree N.
 \Ensure  $NTT\left(P_m \right)=\left(b_0, b_1, \cdots, b_{N-1} \right)$
\State Initializes $b_i = a_i, i \in \left[0, N-1 \right]$; $\omega' = \omega$
\State \textbf{for}{($n=N$, $n>1$, $n=n\gg1$)}  \hfill $\triangleright$ Stage index
\State \quad $\omega_n = \omega^{\frac{N}{n}}$ mod $q$
\State \quad \textbf{for}{($i=0$, $i<n$, $i=i\+n$)} 
\State \quad \quad \textbf{for}{($j=0$, $j<n$, $j\+\+$)} 
\State \quad \quad \quad $b[k\+j]=(b[k\+j] \+ \omega'b[k\+j\+n\text{/}2])$ mod $q$
\State \quad \quad \quad $b[k\+j+n\text{/}2]=(b[k\+j]-\omega'b[k\+j\+n\text{/}2])$ mod $q$
\State \quad \quad \quad $\omega' = \omega \cdot \omega_n$ mod $q$
\end{algorithmic}
\end{algorithm}

\begin{table}[!t]\centering
\caption{Comparison (Radix-2 CT-NTT vs. MAT-based NTT on TPUv4). Latency of 128-batch NTTs (unit: $\mu s$, $N=R\times C$).}\label{tab:ntt_butterfly_vs_recursive}
\vspace{-2mm}
\scriptsize
\begin{tabular}{ccccccc}\hline
Degree ($N$) &$R$ &$C$ &Radix-2 CT-NTT &MAT NTT &Speedup \\\hline
$2^{12}$ &128 &64 &2420 &91.8 &\textbf{26.39}$\times$ \\
$2^{13}$ &128 &64 &4999 &165.4 &\textbf{30.23}$\times$ \\
$2^{14}$ &128 &128 &10530 &355.5 &\textbf{29.5}$\times$ \\
$2^{15}$ &256 &128 &22228 &812.3&\textbf{27.24}$\times$ \\
$2^{16}$ &256 &128 &46996 &1844.8 &\textbf{25.47}$\times$ \\
\hline
\end{tabular}
\vspace{-5mm}
\end{table}
\subsubsection{Basis Conversion (\textbf{BConv})}
\label{sec:basis_change_content}
BConv is used to scale up or down polynomials in HE-Mult and Rotate. For instance, the basis conversion from $\mathcal{B}_1$ (moduli: $q_i$, $i\in[0,L)$) to $\mathcal{B}_2$ (target moduli: $p_j$, $j\in[0,L')$)~\cite{cheon2019full} is expressed as:
\label{sec:bc_two_steps}
\begin{equation*}
\begin{split}
\label{equ:basis_change}
    Conv_{\mathcal{B}_1\rightarrow\mathcal{B}_2} (\mathbf{a}) = 
    ( \Sigma_{i=0}^{L-1} [ a_{n,i} \cdot \hat q_i^{-1}]_{q_i} \cdot [ q_i^{*}]_{p_j} \text{mod } p_j)_{0\leq j < L', 0 \leq n < N}
\end{split}
\end{equation*}
where $\hat q_i^{-1}$ and $[ q_i^{*}]_{p_j}$ could be generated offline and get loaded into the on-chip memory as static parameters during runtime. 

\label{sec:step2}
The above equation is broken into two steps (\figref{fig:bc_two_steps}).

\squishlist
\item Step 1: $b_{n,i} = [a_{n,i} \cdot \hat q_i^{-1}]_{q_i},  0\leq i < L, 0 \leq n < N$ invokes $L$ independent instances of 
$N$-length Vectorized Modular Multiplication, noted as $L\times N$-\texttt{VecModMul} for simplicity.
\item Step 2: $c_{n,j}=\Sigma_{i=0}^{L-1} b_{n,i}\cdot [ q_i^{*}]_{p_j}\ mod\ p_j, 0\leq j < L', 0 \leq n < N$. It invokes one $M_{N\times L'}=M_{N\times L}\cdot M_{L\times L'}$ Modular Matrix Multiplication, noted as $(N,L,L')$-\texttt{MatModMul}
\squishend

\subsection{Final Barrett Modular Reduction in \compiler}
\label{sec:enable_high_precision_mod}

\compiler adopts lazy reduction to allow data to be temporarily over-precision during a chain of compute and defer the final modular reduction to the final result. Specifically, the Montgomery Reduction in Alg. \ref{alg:montgomery_reduction} cannot be used as final modular reduction because its output range is $[0, 2\cdot q_i)$ instead of the desired range $[0,q_i)$. In \compiler, the final modular reduction is achieved through Barrett modular reduction~\cite{Barrett}. 

Specifically, Barrett converts modular multiplication into two multiplications, one shifted multiplication, up-to two subtractions listed in Alg. \ref{alg:barrett} with precision listed in comments. 

\begin{algorithm}
\caption{Modular Multiplication and Barrett Reduction} \label{alg:barrett}
\begin{algorithmic}[1]
 \Require $a,b,m \in \mathbb{Z}_q$, $s = 2\lceil \log_2 q \rceil$, $m=\lfloor 2^s \text{/} q \rfloor$.
 \Ensure  $z=a\cdot b \text{ mod } q$
\State  $z\leftarrow a\cdot b$  \hfill $\triangleright$$\log_2{q}\times \log_2{q}\rightarrow 2\log_2{q}$
\State  $t\leftarrow \left(z\cdot m \right) \gg s$   \hfill $\triangleright$ $2\log_2{q}\times \log_2{q} \gg 2\log_2{q} \rightarrow \log_2{q}$
\State  $z\leftarrow z - \left(t\cdot q\right)$ \hfill  \hfill $\triangleright$ $2\log_2{q} - \log_2{q}\times \log_2{q} \rightarrow 2\log_2{q}$
\State \textbf{if}  {$z\geq q$} 
\State \quad $z\leftarrow z - q$
\end{algorithmic}
\end{algorithm}

\subsection{Fall-back Algorithm for unknown parameters of Fig. \ref{fig:CROSS_Scalar_Flow}}
When all input operands in 32-bit integer arithmetic are not preknown, such as multiplying two input values, BAT is no longer applicable. In this case, \compiler falls back to schedule low-precision chunk-wise multiplications as 1D convolution, as shown in \figref{fig:1d_conv}.

\insertFigure{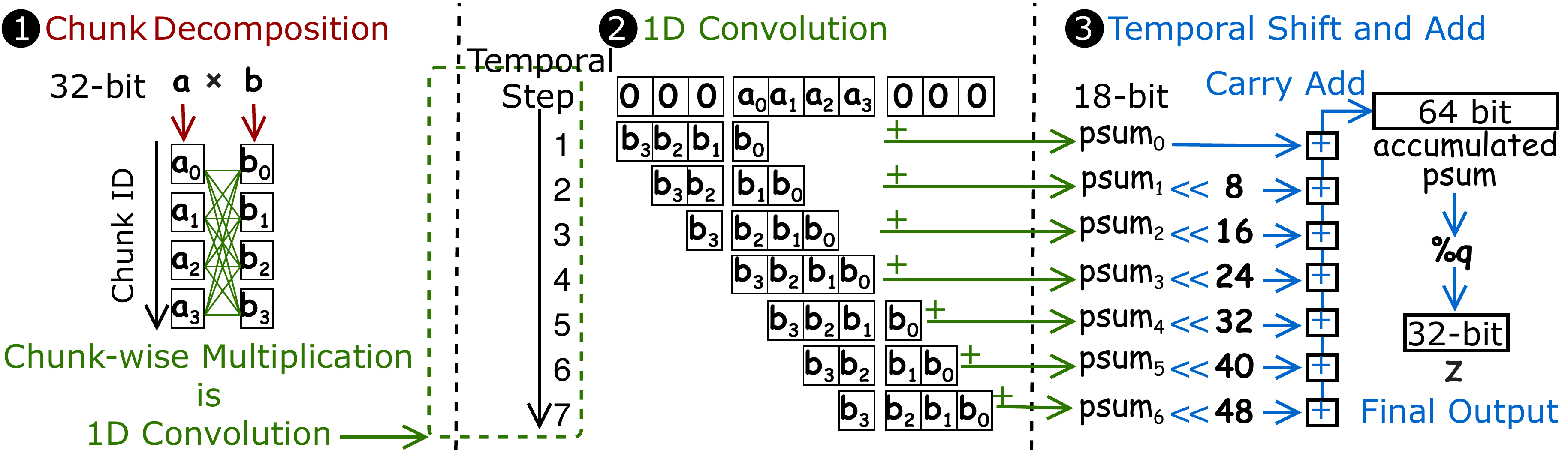}{\compiler maps \textit{high-precision scalar multiplication} into 1D convolution and temporal shifted accumulation when input operands are not known a priori.}

Taking mapping $32$-bit coefficient multiplication to $bp=8$-bit compute as an example, as shown by \ding{182} in \figref{fig:1d_conv}, \compiler segments each coefficient into four uint8 chunks and then maps chunk-wise multiplication as 1D convolution (\ding{183}). Specifically, CROSS reads uint32 from local register, views it as a vector of four uint8 chunks, and pads $\frac{32}{bp}-1=3$ zeros on both sides. Then, convolving the padded $a$ with chunk-decomposed $b$ over seven temporal cycles yields seven partial sums ($psum^{k}, k \in [0, 6]$), each at most $2bp+\log_2\frac{32}{bp}=16+2=18$ bits\footnote{Each chunk-wise multiplication generates 16 bits. A reduction of $\frac{32}{bp}=4$ chunks needs at most 2 extra bits to avoid precision overflow.}. These partial sums are shifted and accumulated (\ding{184}) to $64$-bit final result $psum$, which is stored in eight uint8 registers. $psum$ is finally modular reduced to 32 bits via Barrett Reduction (Alg.~\ref{alg:barrett}). This achieves the same efficiency as sparse matrix multiplication used in TensorFHE(+)~\cite{fan2022tensorfhe}. This algorithm is not used for HE operators, and we detail it here to ensure the general applicability of CROSS.

\subsection{Applying BAT to Preknown Parameters}
We provide Alg.~\ref{alg:HP_on_LP_special_new} to formalize the BAT-enhanced data flow depicted in \figref{fig:CROSS_Scalar_Flow}. BAT is applied to NTT twiddle factors and basis conversion coefficients to maximize throughput on the TPU's MXU. We exclude evaluation keys from BAT because their small reduction size ($dnum \times \text{bytes} \approx 12$) leads to severe MXU under-utilization. While inefficient for the TPU's large systolic arrays, such a transformation could benefit devices with smaller fine-grained matrix engine such as GPUs.


\begin{algorithm}
\caption{{High-Precision Scalar Multiplication using BAT.}} \label{alg:HP_on_LP_special_new}
\begin{algorithmic}[1]
\Require modulus $q$; $a,b,z \in \mathbb{Z}_{q}$ have \textbf{$\log_2 q$} bits; hardware supports $bp$-bit arithmetic; $K \leftarrow \lceil \frac{\log_2 q}{bp}\rceil$, indicating number of bytes.

\item[]
\item[]
\hspace{-2em} {\textbf{\textsc{ConstructToeplitz}}{($[a_k]_{0\le k<K}$)}{} $\rightarrow X$}  \hfill $\triangleright$ \ding{182} in \figref{fig:CROSS_Scalar_Flow}.
\State $X \leftarrow \text{zeros}(2K-1, K)$
\State \textbf{for} $j = 0$ \textbf{to} $K-1$:
\State \quad \textbf{for} $i = 0$ \textbf{to} $K-1$:
\State \quad \quad $X[i+j, j] \leftarrow a_j$
\State \textbf{Return} $X$

\item[]
\item[]
\hspace{-2em} {\textbf{\textsc{BAT}}{($X, bp, q$)}{} $\rightarrow X$} \hfill $\triangleright$ \ding{184} in \figref{fig:CROSS_Scalar_Flow}.
\State \textbf{for} $i = 0$ \textbf{to} $K-2$: \hfill $\triangleright$ Iterate row in bottom block
\State \quad \textbf{for} $j = 0$ \textbf{to} $K-2-i$:
\State \quad \quad $basis \leftarrow (K + i) \cdot bp$
\State \quad \quad $proj \leftarrow (X[i+j+1, j] \ll basis) \bmod q$
\State \quad \quad $[r_k]_{0\le k < 3} \leftarrow \textsc{ChunkDecompose}(proj)$
\State \quad \quad $X[k, K-1-j] \leftarrow X[k, K-1-j] + r_k$, $0\leq k<3$
\State \textbf{Return} $X$

\item[]
\item[]
\hspace{-2em} {\textbf{\textsc{CarryPropagation}}}{($X$, bp)}{} $\rightarrow X$
\item[] \hspace{-2em} $\triangleright$ Ensure all values in $X\leq 2^{bp}-1$.
\State \textbf{for} $j = 0$ \textbf{to} $K-1$:  \hfill $\triangleright$ Iterate all columns
\State \quad \textbf{for} $k = 0$ \textbf{to} (\#Rows of $X$)$- 2$:
\State \quad \quad \textbf{if} $X[k, j] > 2^{bp}-1$:
\State \quad \quad \quad $carry \leftarrow \lfloor X[k, j] / 2^{bp} \rfloor$
\State \quad \quad \quad $X[k, j] \leftarrow X[k, j] \bmod 2^{bp}$
\State \quad \quad \quad $X[k+1, j] \leftarrow X[k+1, j] + carry$
\State \textbf{Return} $X$

\item[]
\item[]
\hspace{-2em} {\textbf{\textsc{OfflineCompile}}{($X, K, bp, q$)}{} $\rightarrow X_{dense}$} \hfill $\triangleright$ \ding{185} in \figref{fig:CROSS_Scalar_Flow}.
\State $\left[a_k\right]_{0\le k < K} \leftarrow$ \Call{ChunkDecompose}{$a$}
\State $X \leftarrow$ \Call{ConstructToeplitz}{$[a_k]_{0\le k < K}$} \hfill $\triangleright$ \ding{182} in \figref{fig:CROSS_Scalar_Flow}.
\State \textbf{while} $(not\ all\ X \leq 2^{bp}-1$ \textbf{or} $X[K:, :] \neq 0)$:
\State \quad  $X \leftarrow \textsc{CarryPropagation}(X)$
\State \quad  \textbf{if} $X[K:, :] \neq 0$ \hfill $\triangleright$ Bottom block contains non-zero.
\State \quad \quad $X \leftarrow \textsc{BAT}(X)$
\State \textbf{Return} $X[0:K, 0:K]$ \hfill $\triangleright$ All values cast to $\text{uint8}$.

\item[]
\item[]
\hspace{-2em} \textbf{\textsc{Main-HPScalarMult}}{($a, b$)} $\rightarrow z$ 
\State $\left[\hat{a}_{i,k}\right]_{0\le i,k < K} \leftarrow$ \Call{OfflineCompile}{$a$} \hfill $\triangleright$ \ding{185} in \figref{fig:CROSS_Scalar_Flow}.
\State $\left[b_k\right]_{0\le k < K} \leftarrow$ \Call{ChunkDecompose}{$b$}
\State $\left[c_{k}\right]$ = $\left[\hat{a}_{i,k}\right] \times \left[b_k\right]$, $0 \le i,k < K$
\State \textbf{for} $k=0$ \textbf{to} $K-1$:
\State \quad $z \mathrel{+}= c_{k} \ll (bp*k)$ \hfill $\triangleright$ \ding{186} in \figref{fig:CROSS_Scalar_Flow}.
\State \textbf{Return} $z$
\end{algorithmic}
\end{algorithm}

\subsection{BAT for Modular Reduction}
\vspace{-1mm}
\label{sec:bat_lazy}
The primary objective of the modular reduction after performing 32-bit multiplication into 64- or 48-bit partial sum in \figref{fig:CROSS_Scalar_Flow} is to compress the 64- or 48-bit partial sum into 32-bit value range required by subsequent pipeline stages, i.e., the final value needs to fit in 32 bits but its actual value could be larger than $q$. Therefore, instead of full modular reduction, we implement a lazy partial reduction to bring 64- or 48-bit partial sum to fit 32 bits (but might larger than $q$). This allows us to only target the ``overflow" bits beyond the 32-bit boundary and apply BAT to it, making it a low-precision MatMul.

This BAT-based formulation enables offloading the modular reduction to the matrix engine, replacing a sequence of vectorized shifts and additions. However, we omit this optimization in our final TPU implementation, because the $4 \times 4$ ($K=4$) reduction dimension is insufficient to saturate the TPU’s 128$\times$128 Matrix Unit (MXU). The resulting systolic array under-utilization makes standard vector instructions more efficient. Conversely, this technique is well-suited for architectures with finer-grained tensor engines (e.g., GPUs), where smaller tile granularities can effectively exploit such small matrix multiplications to achieve higher throughput.


Specifically, the detailed math behind BAT lazy reduction is listed below, assuming multiplication of $a$ and $b$ produces a 64-bit $psum$, which is decomposed into $2K{=}8$ bytes $c_j$, $j\in[0,8)$.
\begin{align*}
& z = psum \bmod q = \sum_{j=0}^{j=2K-1} (\underbrace{c_j}_{\text{one byte}} \times 2^{8j}) \bmod q \\
& =\left( \underbrace{\sum_{j=K}^{j=2K-1} (c_j \times 2^{8j})}_{\text{High 32 bits, need reduction}} + \underbrace{\sum_{j=0}^{j=K-1} (c_j \times 2^{8j})}_{\text{Low 32 bits noted as }low}  \right)\bmod q \\
& =\left( \sum_{j=K}^{j=2K-1} (c_j \times \underbrace{(2^{8j} \bmod q)}_{\text{(BAT) precomputed as } LC_j}) + low \right)\bmod q \\
& =\left( \sum_{j=0}^{j=K-1} (c_{j+K} \times \underbrace{LC_j}_{K\text{ Bytes each}}) + low \right)\bmod q \\
& = \left(\sum_{j=0}^{j=K-1}  ( c_{j+K} \times  \sum_{k=0}^{k=K-1} (\underbrace{ LC_{j,k}}_{\text{one byte each}}\times 2^{8k}) )  + low  \right)\bmod q  \\
& = \left( \sum_{k=0}^{k=K-1} \left(\sum_{j=0}^{j=K-1} {c_{j+K}} \times  LC_{j,k}\right) \times 2^{8k} + low  \right) \bmod q 
\end{align*}
\begin{align*}
& =\left( \sum_{k=0}^{K-1}
\begin{bmatrix}
c_{K} \\
c_{K+1} \\
\vdots \\
c_{2K-1}
\end{bmatrix} \times
\underbrace{\begin{bmatrix}
LC_{0,0}  & \cdots & LC_{0,K-1} \\
LC_{1,0}  & \cdots & LC_{1,K-1} \\
\vdots        & \ddots & \vdots        \\
LC_{K-1,0} & \cdots & LC_{K-1,K-1}
\end{bmatrix}}_{ K\times K \text{ 8-bit matrix}} \cdot \begin{bmatrix}
2^0 \\
2^8  \\
\vdots \\
\vdots 
\end{bmatrix} + low \right) \bmod q 
\end{align*}

\end{document}